\documentstyle[graphicx,amssymb]{elsart}

\begin{document}

\begin{frontmatter}

\begin{flushright}
Fermilab-PUB-05-034-E
\end{flushright}

\title{The Muon System of the Run~II D$\O$ Detector}

\collab{
V.M.~Abazov$^{g}$, B.S.~Acharya$^{f}$, G.D.~Alexeev$^{g}$, G.~Alkhazov$^{k}$, V.A.~Anosov$^{g}$, 
B.~Baldin$^{n}$, S.~Banerjee$^{f}$, O.~Bardon$^{r}$, J.F.~Bartlett$^{n}$, M.A.~Baturitsky$^{g,u}$
% ...
}
 
% ...
\address{$^{f}$Tata Institute of Fundamental Research, Mumbai, India
}
\address{$^{g}$Joint Institute for Nuclear Research, Dubna, Russia 
}
% ...
\address{$^{k}$Petersburg Nuclear Physics Institute, St. Petersburg, Russia 
}
% ...
\address{$^{n}$Fermi National Accelerator Laboratory, Batavia, Illinois 60510 
}
% ...
\address{$^{r}$Northeastern University, Boston, Massachusetts 02115
}
% ...
\address{$^{u}$ Visitors from Institute of Nuclear Problems of the Belarusian State University, Minsk, Belarus 
}
% ...
\thanks[cora]{Corresponding author. Tel.: + 630-840-3851; Fax: + 630-840-8886; {\it E-mail address:} denisovd@fnal.gov}

\date {\today}
\newpage

\begin{abstract}
We describe the design, construction and performance of the upgraded D$\O$ muon 
system for Run~II of the Fermilab Tevatron collider. Significant improvements have been made to the major 
subsystems of the D$\O$ muon detector: trigger scintillation counters, tracking detectors, and electronics. The Run~II central muon 
detector has a new scintillation counter system inside the iron toroid
and an improved scintillation counter system outside the iron toroid.
In the forward region, new scintillation 
counter and tracking systems have been installed. Extensive shielding has been added in the forward region. A large fraction of the muon system electronics is also new. 
\end{abstract}
\end{frontmatter}
\newpage

\section{Introduction}

The D$\O$ detector \cite{refabachi} collected approximately 120 pb$^{-1}$ of data during Run~I of the Fermilab 
Tevatron collider from 1992 to 1996. While the Tevatron was shut down for improvements from 1996 to the beginning of Run~II in 2001, 
the D$\O$ collaboration undertook and completed a substantial detector modification program in order to  
exploit fully the potential of the upgraded Fermilab collider including an increase in instantaneous 
luminosity up to 3$\times$10$^{32}$~cm$^{-2}$s$^{-1}$, an increase in center of mass 
energy from 1.8 to 1.96~TeV, and a reduction in bunch spacing from
3.5~$\mu$s to 396~ns.  With improvements to the Tevatron, we expect 
to collect 4$-$8~fb$^{-1}$ of data  by the end of Run~II in 2009. In this paper we describe the design, 
construction and performance of the D$\O$ Run~II muon system.

The muon system of a general-purpose collider detector should provide efficient muon
triggering and identification with extensive solid angle and momentum coverage and low backgrounds.
Stable and reliable performance over many years of operation and radiation hardness in the high
luminosity hadron collider environment are critical for the muon system as well. The Run~II D$\O$ muon system
was designed with the above goals in mind while also taking into account existing technical constraints 
of the Run~I detector \cite{refbrown} and available resources.

\begin{figure}
\begin{center}
\begin{flushleft}
\hskip -1in
\begin{tabular}{p{160mm}}
\includegraphics[width=1.25\textwidth]{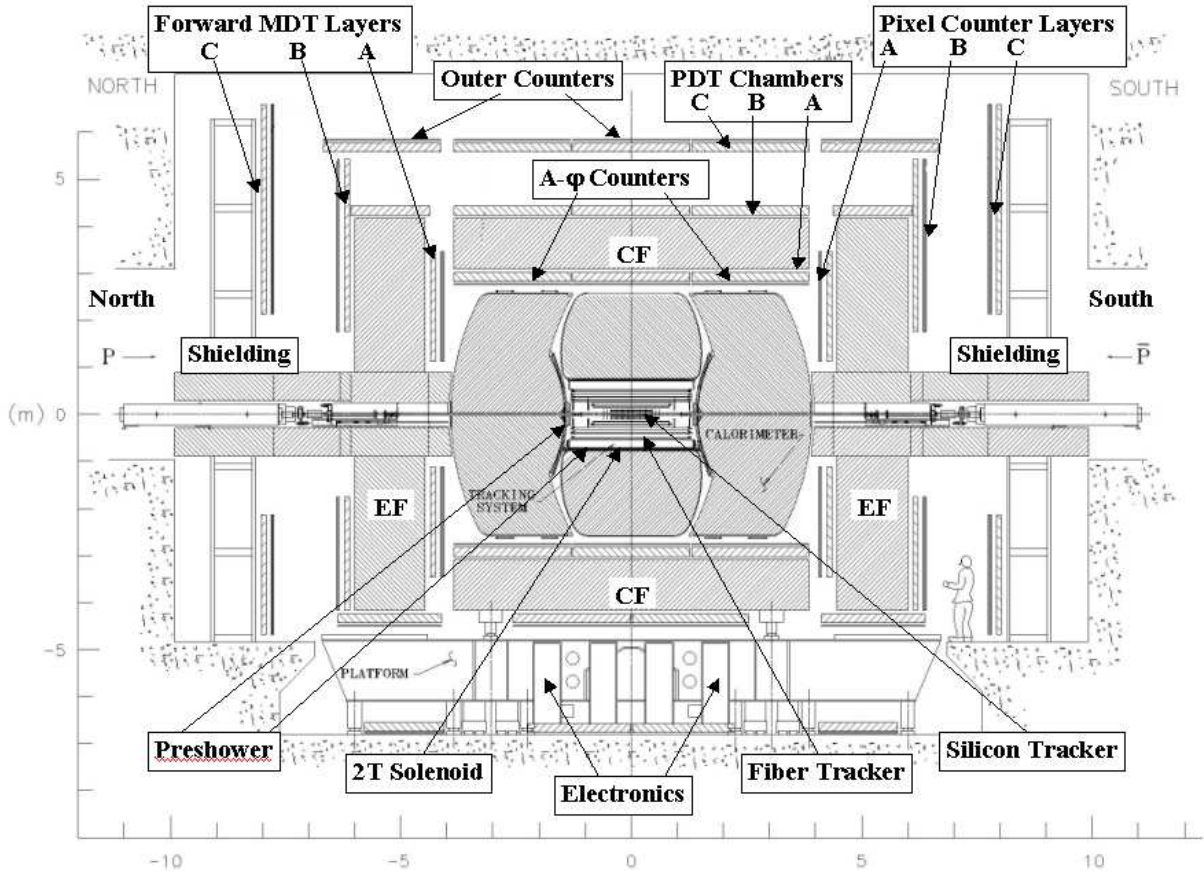}
\end{tabular}
\end{flushleft}
\end{center}
\caption{Cross-sectional view of the D$\O$ Run~II detector.}
\label{picd0}
\end{figure}

The main components of the D$\O$ muon system are identified in the cross-sectional view of the 
Run~II D$\O$ detector shown in Fig.~\ref{picd0}. The D$\O$ coordinate system 
has the ${\it z}$-axis along the Tevatron proton beam direction, ${\it x}$-axis horizontal and pointing out 
of the Tevatron ring, and ${\it y}$-axis pointing straight up. The center of the coordinate system is located 
in the central tracking detector center. 
We use pseudorapidity defined as 
\begin{math}
\eta = -\ln[\tan(\theta/2)],
\end{math}
where $\theta$ is the polar angle measured relative to the ${\it z}$-axis.
The Run~I D$\O$ muon system \cite{refbrown} consisted of two subsystems: the Wide Angle MUon 
System (WAMUS) and the Small Angle MUon System (SAMUS). WAMUS, covering the pseudorapidity 
region $|$$\eta$$|$ $<$ 2.0, consisted of proportional drift tubes (PDTs) 
and three large iron toroidal magnets: a central toroid (CF) and two end toroids (EFs). 
The PDT chambers were arranged in three layers: the A-layer inside the toroids and the B and 
C-layers outside the toroids. SAMUS, covering
2.0 $<$ $|$$\eta$$|$ $<$ 3.0, consisted of a set of drift tube planes and two small 
iron toroids. For Run~II, the PDT
chambers in the forward region (1.0 $<$ $|$$\eta$$|$ $<$ 2.0) were replaced by a new 
tracking system and the SAMUS magnets were
replaced by shielding assemblies. The three main toroids in the D$\O$ muon magnet system, 
the CF and the two EFs, were not changed.
The Run~II D$\O$ muon magnet system is discussed in Section 2. 

Simulation studies and experience from Run~I demonstrated that additional shielding was needed 
to protect the muon system and other components of the D$\O$ detector from the 
increased background expected in Run~II. 
Two sets of shielding assemblies consisting of layers of iron, polyethylene and lead were 
installed in the existing apertures of the end iron toroids and around the accelerator quadrupole magnets. 
The design of the new shielding system is discussed in Section 3.

In Section 4 we discuss three scintillation counter systems added to the original D$\O$ detector. 
Triggering on muons and identifying them in the Run~II environment with shortened bunch spacing 
and increased luminosity requires the use of scintillation counters with 
fine segmentation and good time resolution. The effort to improve the muon trigger began in the latter 
part of Run~I when 240 scintillation counters covering the region $|$$\eta$$|$ $<$ 1.0 
and six of the eight octants in azimuth were installed. 
The solid angle coverage of these counters has now been extended to the bottom of the detector with the
installation of 132 additional counters. This counter system is discussed in Section 4.1. 
In addition to the counters outside of the central iron toroid, two new scintillation 
trigger counter systems have been added for Run~II: a set  
of 630 scintillation counters called A$\varphi$ 
counters are located inside the central toroid and another 4214 counters called ``pixel'' counters are installed in the forward region
(1.0 $<$ $|$$\eta$$|$ $<$ 2.0) \cite{refabramov}. The A$\varphi$ counter system is discussed in Section 4.2 
and the pixel counter system is discussed in Section 4.3.

In Section 5 we describe the muon tracking detector which consists of PDTs in the central rapidity 
region $|$$\eta$$|$ $<$ 1.0 and mini drift tubes (MDTs) in the forward 
region (1.0 $<$ $|$$\eta$$|$ $<$ 2.0). The MDT system is a new tracking system 
that replaces the WAMUS PDT chambers in the forward region.

In Section 6 we discuss the electronics design of the D$\O$ muon system, 
describe methods of electronics synchronization to the accelerator, and present the parameters of the major 
electronics modules designed for
Run~II. A summary is given in Section 7.

\section{Muon toroidal magnets}

\begin{figure}
\begin{center}
\vskip -2cm
\begin{flushleft}
\hskip -1in
\begin{tabular}{p{150mm}}
\includegraphics[width=1.35\textwidth]{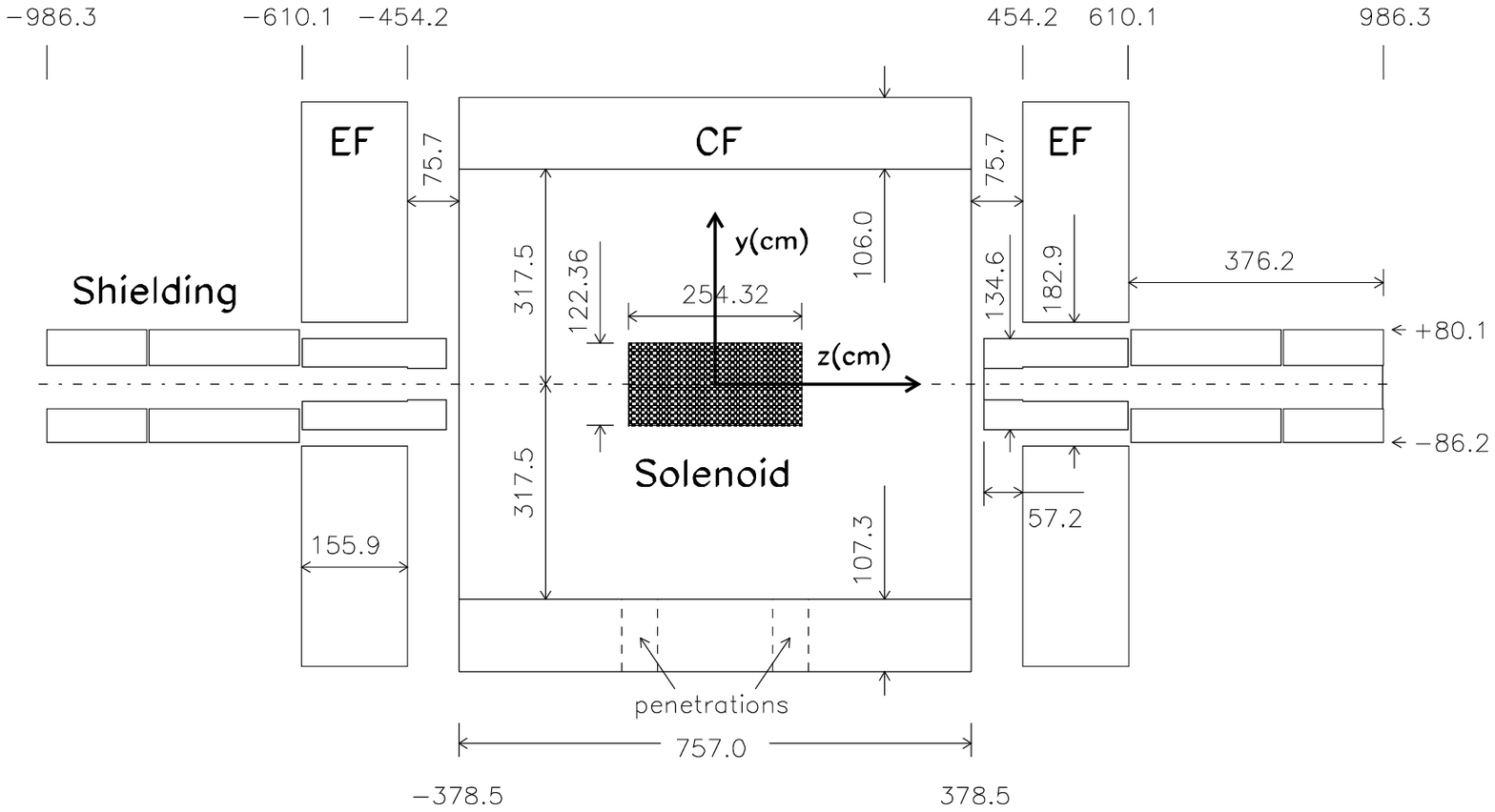}
\end{tabular}
\end{flushleft}
\end{center}
\vskip -4.5cm
\caption{Cross-sectional view of the magnet system and the two shielding assemblies.} 
\label{picmagn}
\end{figure}

\begin{figure}
\begin{center}
\vskip -10cm
\begin{flushleft}
\hskip -1.2in
\begin{tabular}{p{160mm}}
\includegraphics[width=1.5\textwidth]{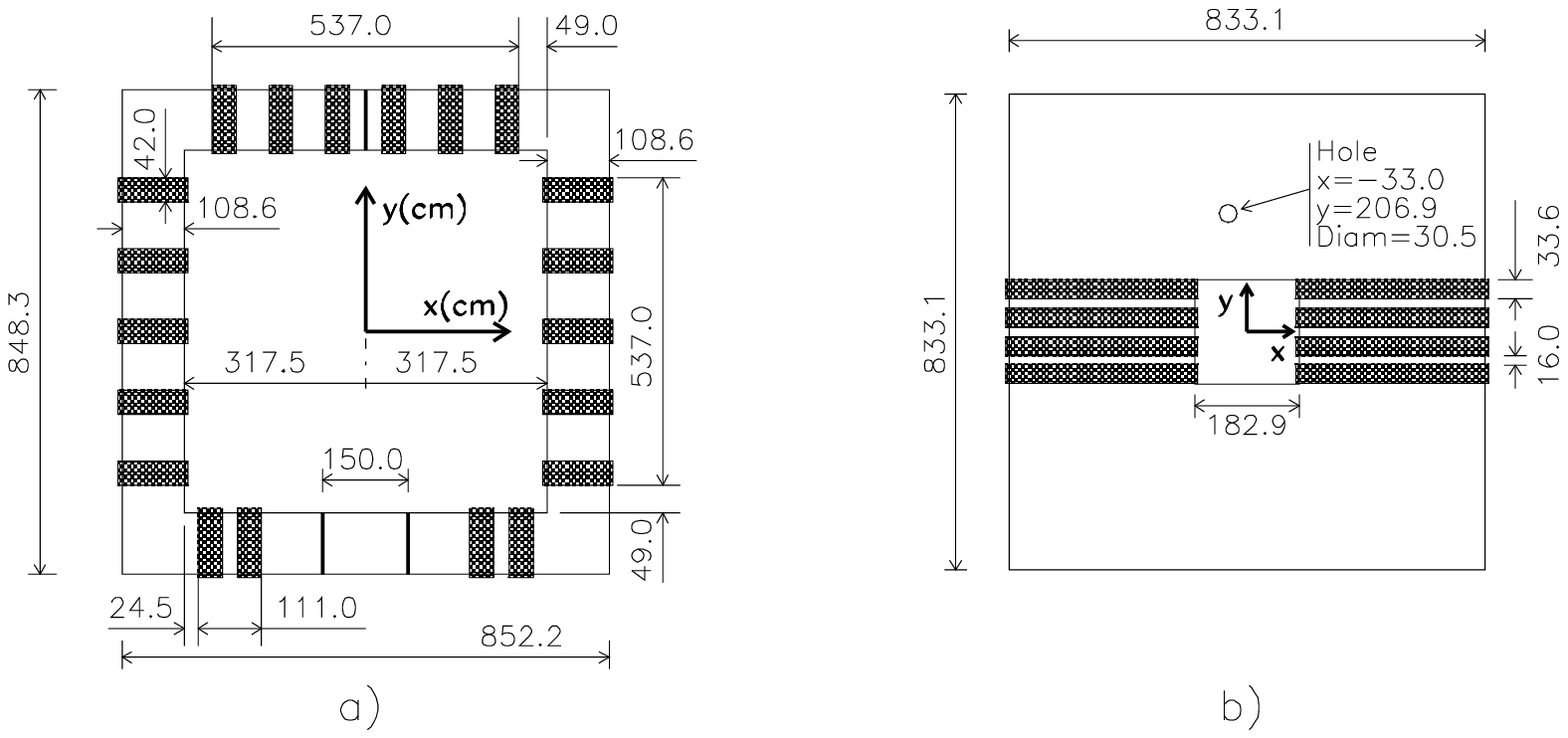}
\end{tabular}
\end{flushleft}
\end{center}
\vskip -1cm
\caption{a) End view of the CF magnet; b) End view of the EF magnet.
The hole at ${\it x}=-33$~cm, ${\it y}=206.9$~cm is for bypass of the Run~I accelerator beam pipe.}  
\label{pictor}
\end{figure}

The three magnets of the Run~II muon detector magnet system \cite{refyamada}; 
the central iron toroid and the two end iron toroids; 
are mounted on the platform that moves the entire detector between detector assembly hall
and the collision hall. 
These toroids account 
for about 65\% of the total 5500 ton weight of the detector.
A cross-sectional view of the Run~II D$\O$ magnet system is presented in Fig.~\ref{picmagn}. End views of 
the CF and EF magnets are shown in Fig.~\ref{pictor}. 
Two small toroids that were located inside the square apertures of the two end toroids as part of 
SAMUS in Run~I were removed and replaced by two shielding assemblies 
that are discussed in the next section.

As indicated in the end view of the CF toroid in Fig.~\ref{pictor}, the CF toroid yoke is made of two side 
pieces of steel plus a central segment at the bottom. The separating lines of the three pieces 
are indicated by the thick solid vertical lines (one at the top and two at the bottom of the sketch of the CF). 
The central segment of the CF is mounted on the detector platform. The calorimeter, the superconducting 
solenoid, and the central tracking systems are supported by this central segment. The two side toroid segments 
move laterally when the detector
is opened to permit access to the inner detector systems. Stainless steel shim plates, 0.48~cm thick 
at the top parting line and 0.32~cm thick at the two bottom parting lines, prevent the three 
magnet elements from ``bonding''
together magnetically due to residual induction in the steel. Penetrations are provided through the 
lower sections of the CF yoke for the passage of the cables of the tracking system. 
The yokes of the EFs are made of steel with central openings.

The central toroid has twenty coils; each end toroid has eight coils. 
Each CF coil has ten turns and each EF coil has eight turns. All coils in the toroidal magnets are connected 
in series and powered by a power supply with a rating of 2500~A at 200~V. 
This power supply is connected in series with a 0.83~m$\Omega$
choke and a reversing switch. The operating current of the toroid coils in Run~I was 2500 A; it 
has been reduced to 1500 A in Run~II providing substantial operational cost savings. As a result, the average magnetic field strength in the toroidal magnets 
is reduced by approximately 6\%.
Since precision muon momentum measurement in Run~II is based on information from the central tracker, 
this minor reduction in magnetic field strength does not adversely affect muon triggering or reconstruction.

\begin{figure}
\begin{center}
\includegraphics[width=1.0\textwidth]{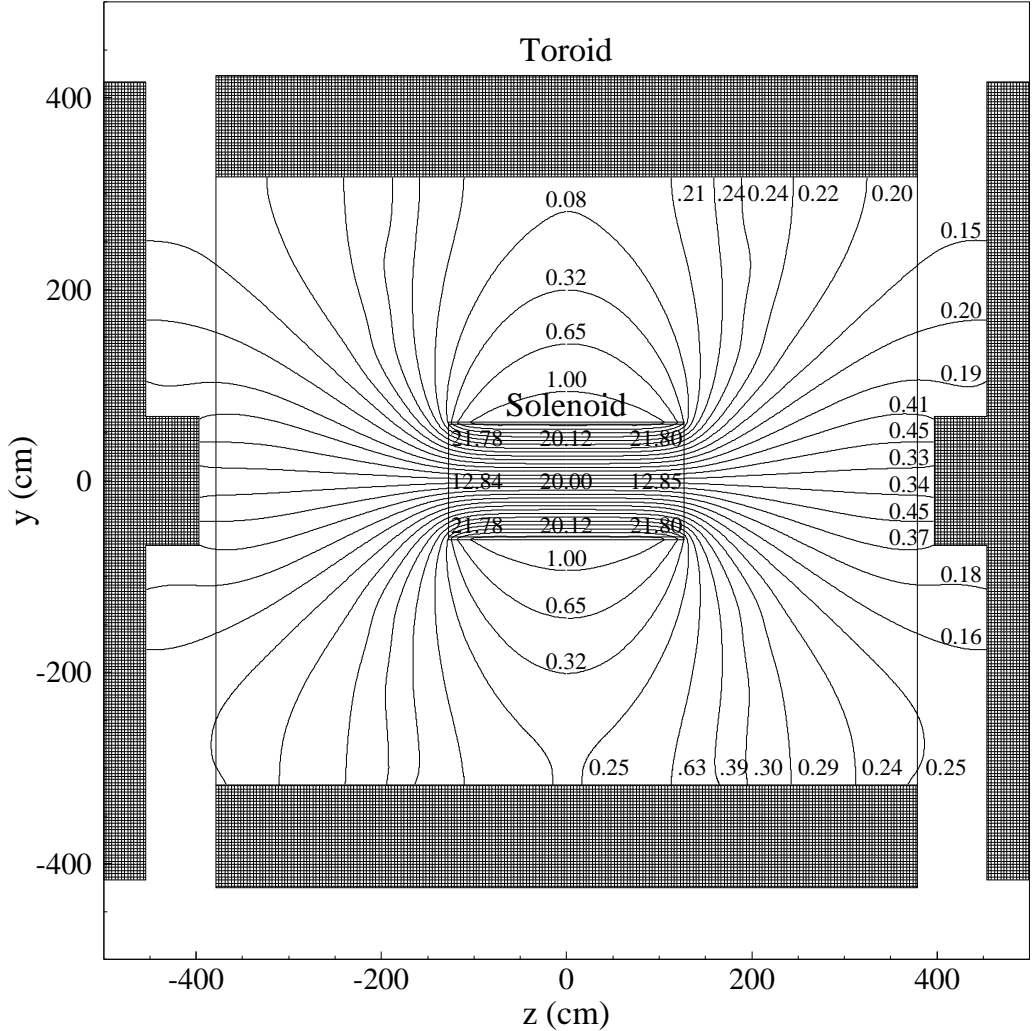}
\end{center}
\caption{Solenoid magnetic field in the ${\it y-z}$ plane of the detector. The magnetic field is in kG.}
\label{picsol}
\end{figure}

\begin{figure}
\begin{center}
\includegraphics[width=1.1\textwidth]{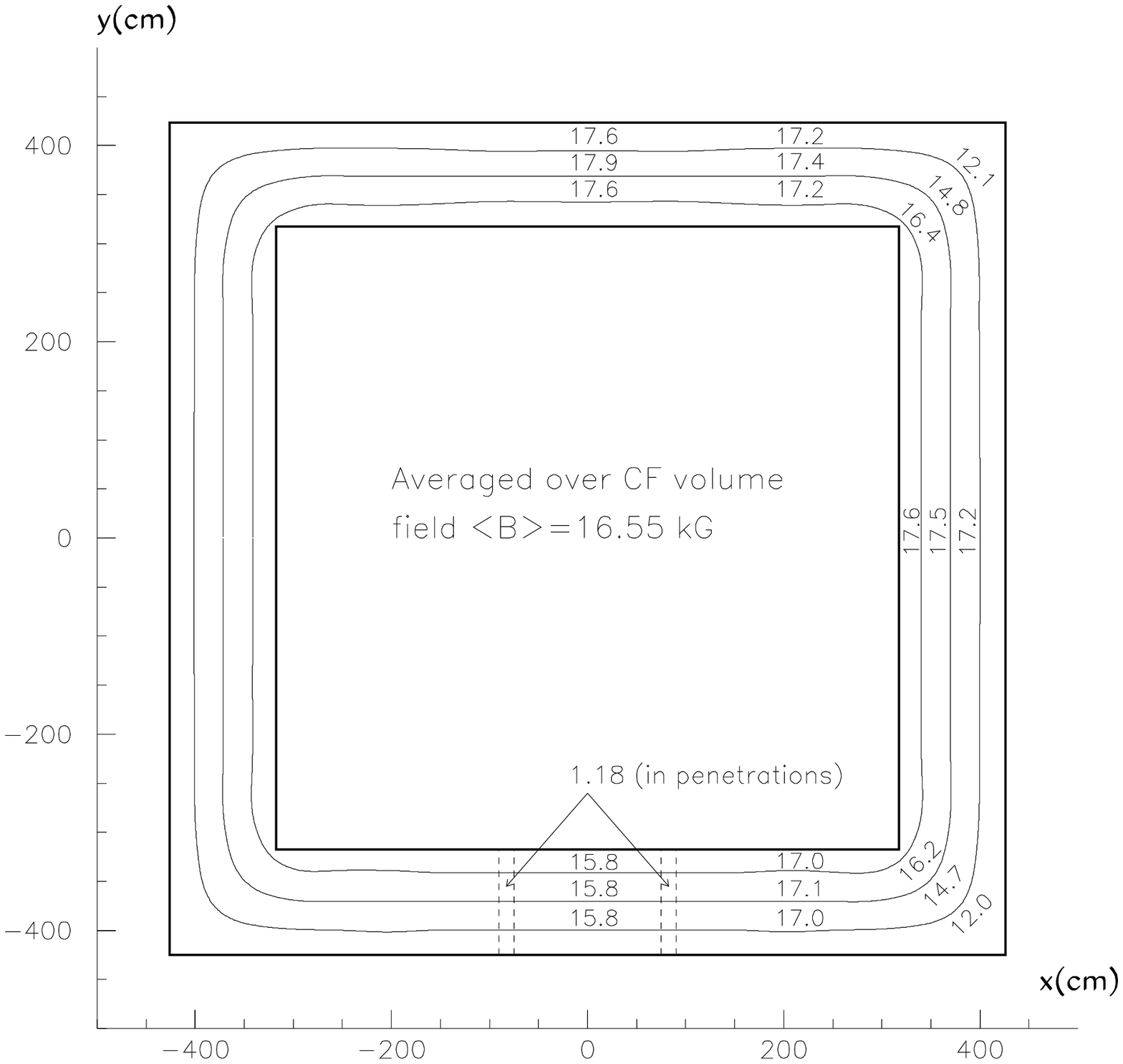}
\end{center}
\vskip -1cm
\caption{Magnetic field in the central toroid magnet. The magnetic field is in kG.}
\label{piccentor}
\end{figure}

\begin{figure}
\begin{center}
\includegraphics[width=1.1\textwidth]{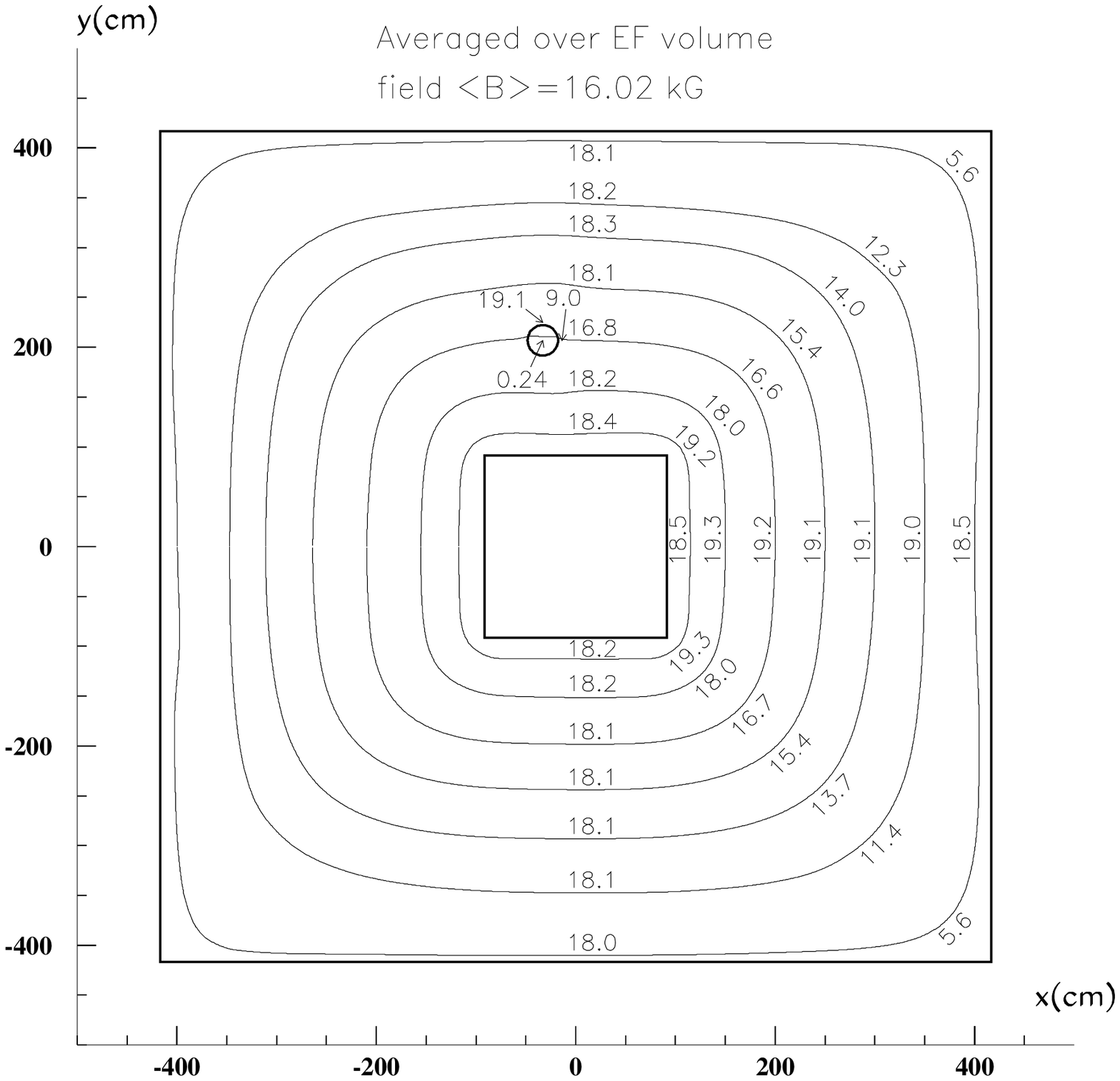}
\end{center}
\vskip -1cm
\caption{Magnetic field in the end toroid magnet. The magnetic field is in kG.} 
\label{picendtor}
\end{figure}

As a major upgrade to the D$\O$ magnet system for Run~II, a superconducting solenoidal magnet with a 2~T magnetic field 
was added inside the inner cavity of the calorimeter. The diameter of the superconducting coil is about 122~cm 
and the total length is about 254~cm. The magnetic field lines of the central solenoid are 
returned by the iron toroidal magnets and by iron in the shielding assemblies. Due to the addition of this 
solenoidal magnet, the magnetic environment of the inner muon detector system and the field distributions 
in the iron toroids are modified. The magnetic field of the Run~II D$\O$ detector magnet system 
is calculated using the program TOSCA \cite{refvector} and corrected using measured values of the field in the steel of the toroids. The estimated 
errors of the field map are 1\% for the iron magnets and 0.1\% for the solenoid. The magnetic 
field associated with the solenoidal magnet is shown in Fig.~\ref{picsol}. The magnetic field inside the CF 
and EF yokes is shown in Figs.~\ref{piccentor} and \ref{picendtor}.

Both toroidal and solenoidal magnets have current reversing switches which are reversed regularly so that 
the experiment collects approximately equal amounts of data in each of the four possible 
field configurations.

\section{\bf Backgrounds and Shielding}

Studies performed during Run~I led us to understand that most hits in the Run~I muon
detectors did not originate from colliding-beam-produced muons. Rather, there were several sources 
of particles that accounted for the high detector occupancies observed. 
First, associated with proton and 
antiproton beams losses in the Tevatron and their subsequent interaction with the accelerator components, 
there was a background of particles coming from the accelerator tunnel into the D$\O$ collision hall. 
To suppress this background, two protective measures have been developed based on MARS \cite{refmokhov} 
calculations and implemented during the 1994 Tevatron shutdown \cite{refbutler}: (i) two 
2-m-thick concrete walls were installed in the tunnel on both sides of the D$\O$ collision 
hall; (ii) the accelerator A0 scraper was replaced with a new one with two thin scattering targets. 
After these steps, the Tevatron halo backgrounds became negligible compared to other 
backgrounds \cite{refbutler,refdiehl}. 
Second, the remnants of the proton and antiproton, after their collision at the interaction point, 
interact with the beam pipe, the edges of the forward calorimeter near the beam pipe, and the 
accelerator's low-beta quadrupole magnets.  
The background from the forward calorimeter edges was 
seen in all muon detector components inside the toroid, including the central muon detectors. The time 
spectrum of these particles was measured with scintillation counters. The arrival time was as expected 
for particles originating near the beam pipe at the calorimeter edges. Because the path was longer than 
that of muons originating from collisions, the background particles arrive measurably later than 
muons originating in the detector center. 
The background from the low-beta quadrupole
magnets was detected in all parts of the forward muon detector and in those parts of the central muon 
system not protected by the EF magnets. Cosmic ray muons are only 
a small contributor to the detectors' occupancy.

\begin{figure}
\begin{center}
\includegraphics[width=1.1\textwidth,height=0.77\textwidth]{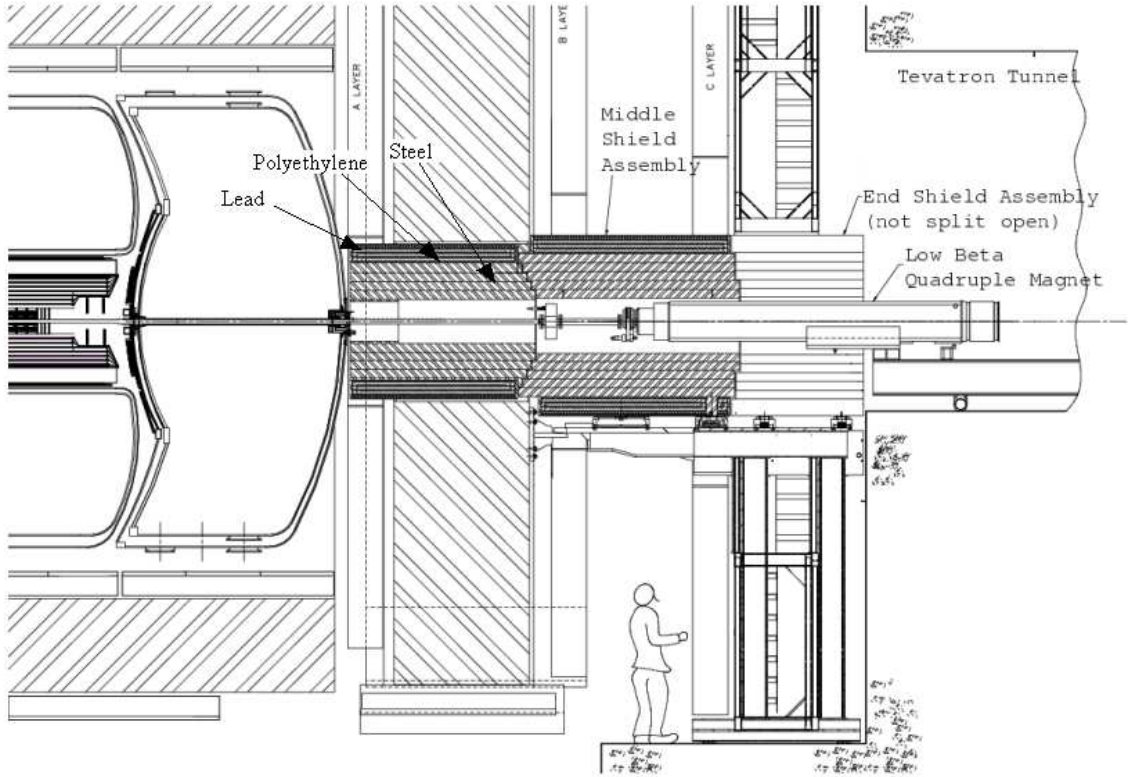}
\end{center}
\caption{Shielding assembly.}
\label{picshield}
\end{figure}

In response to the experience in Run~I, we have performed a detailed analysis of shielding for the 
Run~II muon detector \cite{refdiehl,refsirotenko}. Simulations were performed with GEANT \cite{refcosmo} and 
MARS Monte Carlo codes. The new shielding design includes two 100-ton assemblies surrounding 
the beam pipe and the low-beta quadrupole magnets on both ends of the detector. The shielding assembly 
extends from the end of the calorimeter to the wall of the collision hall. To permit articulation 
of the detector during opening and roll-in, each of the two shielding assemblies is made in three 
overlapping segments. Fig.~\ref{picshield} shows a shielding assembly in the detector operating position.
During the detector opening process, 
the end section of the shielding assembly is first split open and the middle section is moved away from detector center
on its support rails. The front section of the shielding assembly is supported by the EF toroid and moves together with the EF toroid.

The shielding assemblies consist of layers of iron, polyethylene, and lead. 
The innermost layer is made of iron because it has reasonably short interaction (16.8~cm) 
and radiation (1.76~cm) lengths, and is relatively inexpensive. The optimal thickness of the individual 
layers is determined by Monte Carlo simulations (GEANT and MARS).
The iron absorbs electromagnetic and hadronic showers that make up the majority of the background. 
However, iron is almost transparent to low energy neutrons. Polyethylene is used as a neutron absorber 
because it is rich in hydrogen and effective in moderating neutrons.  Neutron capture in the polyethylene 
results in gamma rays. They are, in turn, absorbed by the outermost lead layer. The shielding is 
a square tube with wall thickness of 51~cm of iron followed by 15~cm of polyethylene and 5~cm of lead. 

\begin{figure}
\begin{center}
\includegraphics[width=1.0\textwidth,height=1.0\textwidth]{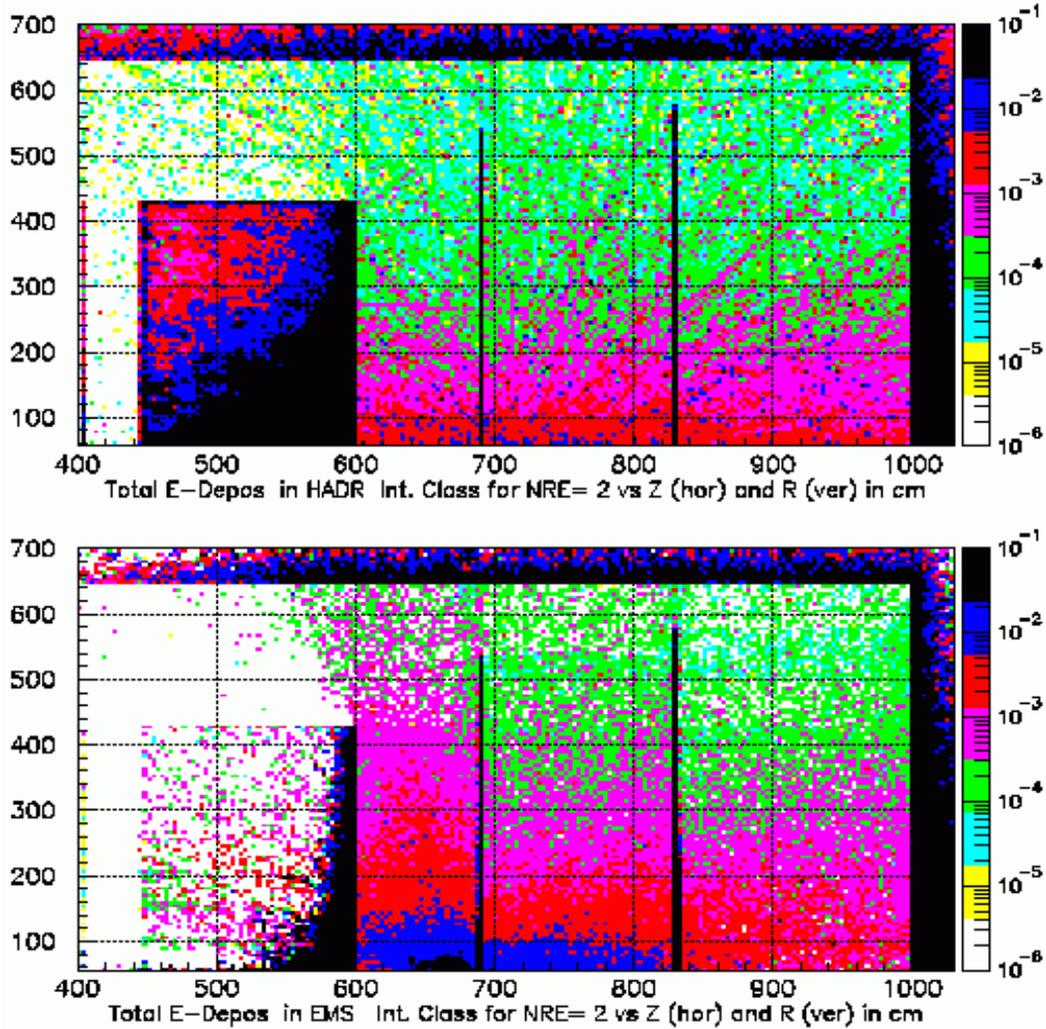}
\end{center}
\caption{Energy depositions from hadron (top) and electromagnetic showers (bottom) in the D$\O$ muon 
detector without the shielding system. The units of the energy deposition scale shown on the right side of the 
figure are GeV/cm$^3$ per sec
for an instantaneous luminosity of
2$\times$10$^{32}$~cm$^{-2}$s$^{-1}$.
The darker shades indicate higher energy deposition. Vertical axis is
${\it y}$ and horizontal axis is ${\it z}$ coordinate in the D$\O$ coordinate system given in cm.}
\label{picener1}
\end{figure}

\begin{figure}
\begin{center}
\includegraphics[width=1.0\textwidth,height=1.0\textwidth]{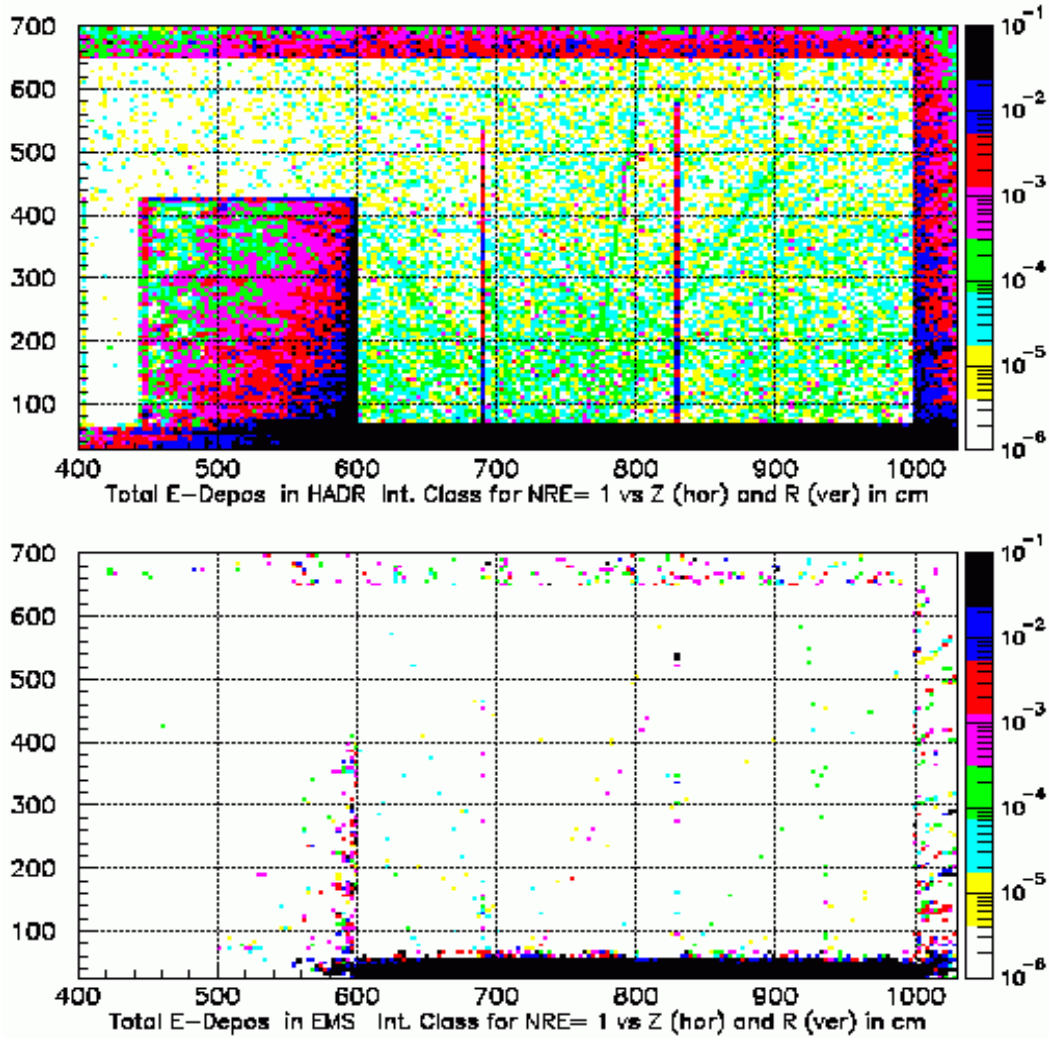}
\end{center}
\caption{Energy depositions from hadron (top) and electromagnetic showers (bottom) in the D$\O$ muon 
detector with the shielding system. The units of the energy deposition scale shown on the right side of the 
figure are GeV/cm$^3$ per sec
for an instantaneous luminosity of
2$\times$10$^{32}$~cm$^{-2}$s$^{-1}$.
The darker shades indicate higher energy deposition. Vertical axis is
${\it y}$ and horizontal axis is ${\it z}$ coordinate in the D$\O$ coordinate system given in cm.}
\label{picener2}
\end{figure}

The comparison between simulated energy depositions from hadron and electromagnetic 
showers at an instantaneous luminosity of
2$\times$10$^{32}$~cm$^{-2}$s$^{-1}$ can be seen in Figs.~\ref{picener1} and \ref{picener2}.
Fig.~\ref{picener1} shows the muon system without and Fig.~\ref{picener2} with the shielding system. 
The three scintillation counter layers in the detector forward region 
(seen as vertical lines in the figures) are located approximately 400, 690 and 830~cm from 
the interaction region. The grayscale of the detector areas indicates the level of the radiation dose.  
Using the Monte Carlo simulation, we find that the energy deposition in the scintillator planes is 
reduced by a factor of approximately 10$^2$. As a result, the number of background hits in the 
muon counters is reduced 
by a factor of between 40 and 10$^2$ \cite{refdiehl,refsirotenko}. Background rates obtained during Run~II agree 
with the Monte Carlo simulation within 50\%. For example, occupancy of the B and C-layers of the forward 
trigger scintillation counters (Section 4.3) for minimum bias event at a luminosity 
of 2$\times$10$^{31}$~cm$^{-2}$s$^{-1}$ is 0.02\%.

\section{Scintillation trigger counters}

The triggering capability of the Run~II D$\O$ muon system has been greatly enhanced 
by the addition of scintillation counters with fine segmentation and good time resolution. The central 
region ($|$$\eta$$|$ $<$ 1.0) of the Run~II D$\O$ detector is covered by two layers of scintillation counters, 
one layer outside of the central toroid and one layer inside. The forward regions
(1.0 $<$ $|$$\eta$$|$ $<$ 2.0) are covered by three layers of scintillation counters, 
one layer inside the end toroid and two layers outside. These counters are used to trigger on muons 
associated with proton-antiproton collisions and to provide precise timing information for 
reconstructing muon trajectories using the muon tracking chambers. Fine segmentation and good timing 
resolution of these counters are essential for rejecting cosmic ray muons and other out-of-time backgrounds 
discussed previously.  

The design {\it p\/}{\it $_{T}$} threshold for the unprescaled muon triggers in Run~II 
is 6~GeV/$c$ for a single muon trigger and 3~GeV/$c$ for a dimuon trigger. The Level 1 muon trigger combines 
track candidates in the central fiber tracker \cite{refgaston} with information in the muon scintillation 
counter system. Therefore, the layout of the scintillation counters is required to match 
the 4.5$^{\circ}$ trigger segmentation in $\varphi$ of the fiber tracker. 
More information about the D$\O$ three-level trigger system is given in Section 6.

\subsection{Scintillation counters outside the central toroid}
 
During the latter part of Run~I, 240 scintillation counters for muon triggering were 
installed on the top and two sides of the C-layer PDT chambers. 
Details of this counter system are given in Ref.~\cite{refacharya}. These scintillation counters 
span $|$$\eta$$|$ $<$ 1.0 and cover six octants (top and sides) in azimuth. 

For Run~II, an additional 132 scintillation counters covering the two bottom octants were 
installed under the central toroid and the two end toroids. The design of these counters is similar 
to the counters installed previously. They are made of scintillator sheets with embedded wavelength shifting 
(WLS) fibers for light collection. EMI 9902K \cite{refemi} photomultiplier tubes 38~mm in diameter are 
used to read out the light produced by the counters. The size of these counters is 
approximately 200~cm $\times$ 40~cm. They are arranged in two layers (B and C) and are oriented with 
the short dimension (40~cm) along the
$\varphi$ direction, so that each counter covers approximately 4.5$^{\circ}$ in $\varphi$.  

\subsection{ A{\it $\varphi$} scintillation counters inside the central toroid}

The A{\it $\varphi$} counter system added for Run~II has 630 counters mounted on the A-layer 
PDT chambers between the central iron toroid and the calorimeter covering $|$$\eta$$|$ $<$ 1.0. 
Each counter approximately matches the 4.5$^{\circ}$ {\it $\varphi$}-segmentation of the central fiber tracker. 
There are nine counter barrels located within 
\begin{math}
-1.0<\eta<1.0. 
\end{math}
Combining the ``in-time'' hits from the A{\it $\varphi$} counters with tracks in the central fiber 
tracker is a key element of muon triggers with low {\it  p}{\it $_{T}$} thresholds. 
The time gate for A{\it $\varphi$} counters signals is 24~ns at the Level 1 trigger.  
In addition, the A{\it $\varphi$} counters provide timestamps to reconstruct muon tracks in the A-layer PDT chambers. 
This is particularly important for low-{\it p}{\it $_{T}$} muons that do not penetrate the toroidal magnet. 

\subsubsection{Counter design}

\begin{figure}
\begin{center}
\includegraphics[width=1.0\textwidth,height=0.85\textwidth]{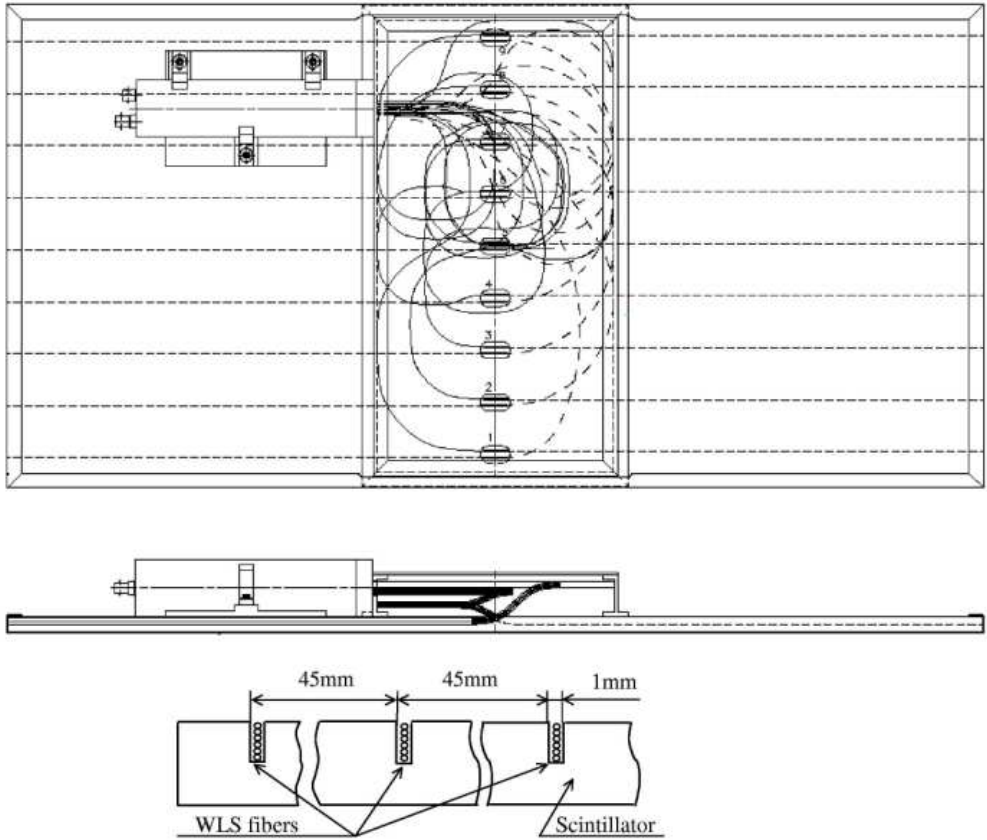}
\end{center}
\caption{Top and side views of an A$\varphi$ counter and counter cross section.}
\label{picaphi1}
\end{figure}

The A{$\varphi$} counters are made from 12.7-mm-thick BICRON 404A scintillator with BICRON BCF 92 
wavelength shifter fiber \cite{refbicron} embedded
in machined grooves \cite{refevdoki} (see Fig.~\ref{picaphi1}).
This WLS has a fast (2.7~ns) decay time and its absorption spectrum matches 
the 420~nm emission peak of the 404A scintillator. The grooves are machined 4.5~cm apart and 
6~mm deep and run 
from the center of the counter to its edge. Six individual 
fibers are glued into each groove and tapered out of the groove at the middle of the counter. 
Fibers from 10 to 16 grooves are combined into a single bundle that directs light to a 25-mm-diameter  
green-extended 115M phototube made by MELZ \cite{refjoint}.
 This phototube has an average quantum efficiency of 15\% at 500~nm with a maximum gain 
around 10$^{6}$. The sensitivity peak of the 115M photocathode matches the 480~nm emission peak 
of the WLS fibers. The phototube is secured to the counter case at the center of the
counter. All fibers are within 5~mm of being the same length. Each counter is wrapped 
in a layer of TYVEK \cite{refdupont} type 1056D sheet with a layer of black TEDLAR \cite{refdupont} over 
the TYVEK for light-tightness. 
The counter case is an aluminium box with welded corners. The light-tight case provides mechanical 
protection for the counter, supports the PMT assembly, and provides counter mounts.

The A{\it $\varphi$} counters operate in the 200 to 350~G residual magnetic field of the muon toroidal and 
central solenoidal magnets. Studies of magnetic shielding have been performed \cite{refbez} with the goal 
of reducing the magnetic field in the phototube region to about 1 G, which is necessary for high 
gain operation. 
The shielding consists of a 6-mm-thick soft steel tube and a 1.2-mm-thick $\mu$-metal cylinder. 
With this shielding, the phototubes can operate without significant gain reduction in a 700~G magnetic 
field perpendicular to the phototube axis. For fields parallel to the tube axis, the gain 
reduction is less than 5\% at 250~G and less than 10\% at 350~G.

\subsubsection{Counter arrangement and mounting}

\begin{figure}
\begin{center}
\includegraphics[width=1.0\textwidth,height=0.85\textwidth]{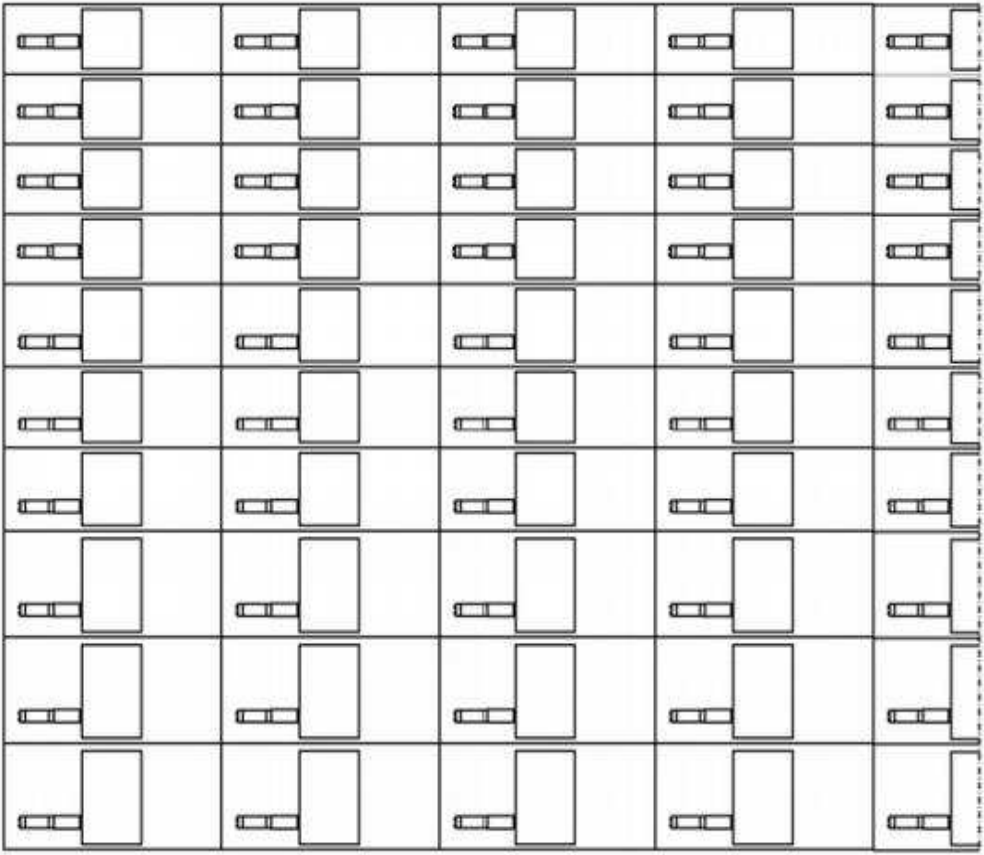}
\end{center}
\caption{The arrangement of A$\varphi$ counters in one of the side quadrants. One quarter of a quadrant (four and one half barrels and 10 columns) is shown.}
\label{picaphi2}
\end{figure}

The A{\it $\varphi$} counters cover the four quadrants of the A-layer PDT chambers. The length of 
the A-layer PDT chambers in the beam direction is approximately 7.6~m.  Nine barrels of counters cover 
the entire span $|$$\eta$$|$ $<$ 1.0 in the beam direction. The length of each A{\it $\varphi$} 
counter is 84.5~cm. Twenty counter columns, of three different widths, cover one quadrant in the 
azimuthal direction. Three different widths; large, medium and small; allow each column of counters 
to cover approximately 4.5$^{\circ}$ in {\it $\varphi$}, matching the {\it $\varphi$}-segmentation of the fiber 
tracker trigger. 
The numbers and dimensions of the three differently sized counters are given in Table~1. Fig.~\ref{picaphi2} shows 
the counter arrangement in a quadrant and
Table~2 summarizes the counters' segmentation.

\begin{table}
\begin{center}
Table 1 \\
Numbers and sizes of A{\it $\varphi$} counters. \\
\vskip 2mm
\begin{tabular}{|c|c|c|} \hline
Number&Length (cm)&Width (cm) \\ \hline
216&84.5&36.7 \\
144&84.5&27.5 \\
270&84.5&23.1 \\ \hline
\end{tabular}
\end{center}
\end{table}

\begin{table}
\begin{center}
Table 2 \\
Segmentation of the A{\it $\varphi$} counters in a quadrant. \\
\vskip 2mm
\begin{tabular}{|l|c|c|c|} \hline
Counter&Top&Side&Bottom \\
Width (cm)&quadrant&quadrants&quadrant \\ \hline
36.7&3&3&3 \\
27.5&1&3&1 \\
23.1&12&8&2 \\
27.5&1&3&1 \\
36.7&3&3&3 \\
Total number&9 $\times$ 20&2 $\times$ 9 $\times$ 20&9 $\times$ 10 \\ \hline
\end{tabular}
\end{center}
\end{table}

The edges between neighboring counters in azimuth overlap to prevent inefficiency. 
The average overlap between counters is about 3\% of the counter width. In the longitudinal direction, 
the counters are mounted end-to-end with a 10~mm gap between scintillators.  
The bottom quadrant is instrumented with only ten counter columns because of the calorimeter support 
and other mechanical obstructions. 
The counters are mounted on aluminium cross members attached to steel brackets that 
are secured to the edges of the PDT chambers. The mounting system of the A{\it $\varphi$} counters ensures  
that the relative positions of the counters are fixed to within about 1 mm. The actual position 
of each counter is measured to an accuracy of 3~mm. 

\subsubsection{Aging of counters}

The A{\it $\varphi$} counters are shielded by the calorimeter and consequently the 
expected Run~II radiation dose is less than 1 krad in the hottest region. 
From studies of the radiation effects on the light output and attenuation length for three different 
types of scintillation materials
and three WLS materials \cite{refbez}, we concluded that the best performance 
in terms of radiation hardness is obtained by using the 
combination of BICRON 404A scintillator and BCF 91A WLS fibers \cite{refbicron}. After 20 krad, the light output from 
counters made from BICRON 404A plus BCF 91A fiber decreases by only 1\% compared to 15\% for 
a counter that uses BCF 92 fiber. This is caused mainly by the decreased attenuation length in the 
WLS fibers due to radiation damage.  
Since the expected dose is less than 1 krad in the highest 
dose location (for total delivered luminosity of 10~fb$^{-1}$), BCF 92 fibers were selected for better 
light output and timing. We expect no significant performance degradation of A{\it $\varphi$} counters 
in Run~II due to radiation damage.

The effect of phototube aging is also expected to be small. The gain reduction of phototubes after 100~C charge 
deposition on the phototube anode is about 10\%. The charge deposition for the phototubes in 
the A$\varphi$ counters is expected to be well below 100~C. Any small gain reduction can be compensated 
for by adjusting the signal discrimination threshold or phototube operating voltage.

\subsubsection{Counter performance}
 
During prototype counter tests the average integrated pulse height was measured to be approximately 55 photoelectrons for cosmic ray muons 
that penetrate the middle of the counter. The measured reduction of signal amplitude for the area that 
is farthest away from the phototube is approximately 7\%. The time resolution of the counter is measured 
to be 0.8~ns, for 20~mV discrimination threshold, while muon signal amplitude is adjusted 
to be approximately 0.10~V.

During counters production, the light yield for each scintillation counter was measured 
using cosmic muons and the gain of each phototube was measured on a test stand. 
Phototube-to-counter matching was then done to obtain 
the same signal amplitudes in each group of fifteen counters connected to the same 
high voltage power supply during detector operation. 

\begin{figure}
\begin{center}
\includegraphics[width=1.1\textwidth]{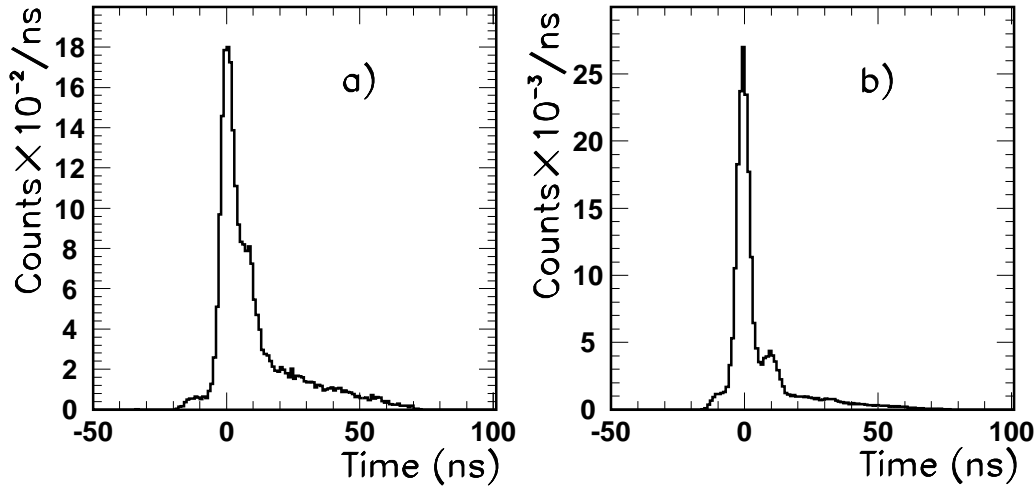}
\end{center}
\vskip -6.5cm
\caption{The time-of-arrival distributions recorded by A$\varphi$ counters  in events 
triggered by a) all triggers; b) muon triggers only.}
\label{picaphitime}
\end{figure}

The performance of the installed A{\it $\varphi$} counter system during Run~II has been monitored using muons 
from proton-antiproton collisions. The time-of-arrival of hits using all D$\O$ 
triggers is shown in Fig.~\ref{picaphitime}a and the time-of-arrival of hits using muon triggers based on 
the central scintillation counters only is shown in Fig.~\ref{picaphitime}b. The peaks around zero in both 
figures are from muons originating at the interaction region. The time-of-arrival distribution 
shown in Fig.~\ref{picaphitime}a 
includes a significant number of late hits from the various background sources discussed in Section 3. 
A secondary peak near 10~ns that is visible in Fig.~\ref{picaphitime}b is caused by backscattered particles 
from the edges of calorimeter. The time resolution of all 630 A{\it $\varphi$} counters combined is 
\begin{math}
\sigma_{t}=2.5 
\end{math} 
ns without corrections for differences in time-of-flight, light 
propagation in the counters or cable length within each group of fifteen counters. Other factors 
that contribute to the measured time resolution of the A{\it $\varphi$} counters include the intrinsic resolution 
of the counter and variations in the ${\it z}$-position of the interaction vertex.

\subsection{Scintillation Trigger Counters in Forward Regions}

\subsubsection{System layout}

\begin{figure}
\begin{center}
\includegraphics[width=1.0\textwidth,height=0.8\textwidth]{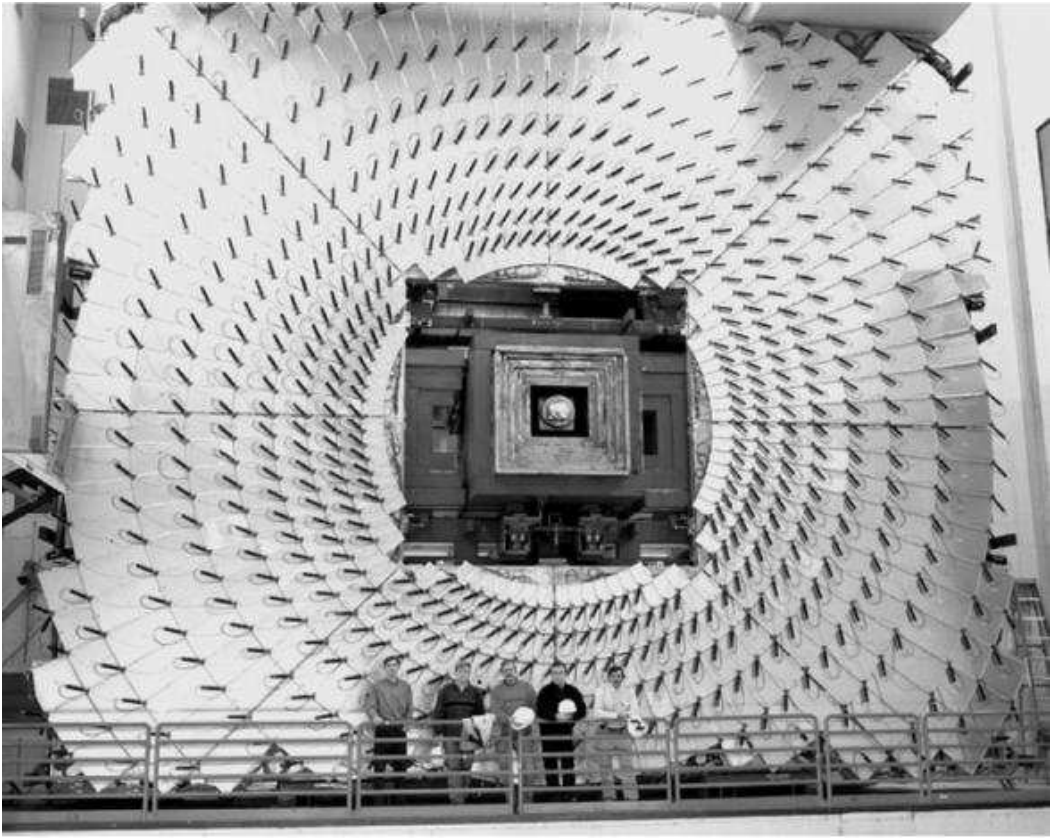}
\end{center}
\caption{A photograph of C-layer pixel counters.}
\label{picpixphoto}
\end{figure}

The ``pixel'' counter system covering 1.0 $<$ $|$$\eta$$|$ $<$ 2.0 consists of 4214 
trapezoidal-shaped scintillation counters. These are arranged so that muons originating at the interaction region 
traverse the three counter layers, A, B and C, shown in Fig.~\ref{picd0}. The A-layers and the B-layers 
are mounted on the inside and outside faces of the two EF toroids respectively, while the C-layers are mounted on two separate
steel structures mounted on sidewalks of the detector hall.  The C-layer, the largest among the three layers, is approximately
12$\times$10~m$^{2}$. A photograph of C-layer pixel counters is shown in Fig.~\ref{picpixphoto} where 
the shielding around the beam pipe described in Section 3 is also clearly seen.

In each layer, the trapezoidal-shaped counters are arranged in an $r$ - $\varphi$ geometry in about 
twelve concentric zones in the radial direction. The $\eta$-segmentation is 0.12 for the nine inner zones 
and 0.07 for the three outer zones. The $\varphi$-segmentation is 4.5$^{\circ}$. The choice of the 
segmentation is driven by the $\varphi$-segmentation of the central 
fiber tracker trigger, the minimum muon trigger momentum threshold, muon multiple scattering in the toroids, 
background trigger rates due to accidental coincidences, and the total number of counters required. 
In the forward region where the background rates are high, the granularity must be sufficiently fine to keep 
the combinatoric background trigger rate low. For three layers of ${\it N}$ counters, the combinatoric background rejection 
scales as ${\it N}$$^{3}$ assuming hits in the layers are uncorrelated, so small changes in granularity have 
a big effect on the combinatoric background.

The size of the smallest counters in the A-layer is 9$\times$14~cm$^{2}$ and the size of the 
largest counters in the C-layer is 60$\times$110~cm$^{2}$. 
The minimum size of the pixel counters is defined by the 3$^{\circ}$  multiple scattering angular 
spread of a ${\it p}$$_{\it{T}}$~=~3~GeV/$c$ muon through the end toroid. As mentioned earlier, triggering on muons down 
to ${\it p}$$_{\it{T}}$ = 3~GeV/$c$ is one of the basic requirements for the D$\O$ muon trigger system.  

\subsubsection{Pixel counter design and mounting}

Several light collection methods were studied in order to optimize the pixel counters for light 
yield, uniformity, time resolution, and background rejection while maintaining a reasonable 
production cost. From measurements \cite{refevdoki}, we concluded that for the geometry of our counters, WLS 
bars give better light collection efficiency and allow a simpler construction procedure than 
WLS fibers. The design using WLS bars for light collection was chosen as the basic option for the 
large majority of the counters \cite{refbez}.  

The counters use Kumarin 30 type WLS bars \cite{refkumarin} for light collection and 12.7-mm-thick 
BICRON 404A scintillator plates of trapezoidal shape. BICRON 404A scintillation light emission peak is 420 nm, 
its decay time is 2.0~ns, and its attenuation length is 1.7~m.  The absorption peak of the Kumarin 30 WLS bar 
matches the emission peak of the scintillator.  The light emission peak of the Kumarin 30 WLS bar is 480~nm, 
the decay time is 2.7~ns, and the attenuation length is 1.4~m.  

Effects of radiation damage on the light output of Kumarin 30 WLS bars and BICRON 404A scintillator 
were studied using sample counters of different sizes \cite{refbez}. We concluded that the expected integrated 
Run~II radiation level of 1 krad for an integrated luminosity of 10~fb$^{-1}$ 
is significantly below what would degrade pixel counter performance.

\begin{figure}
\begin{center}
\includegraphics[width=1.1\textwidth]{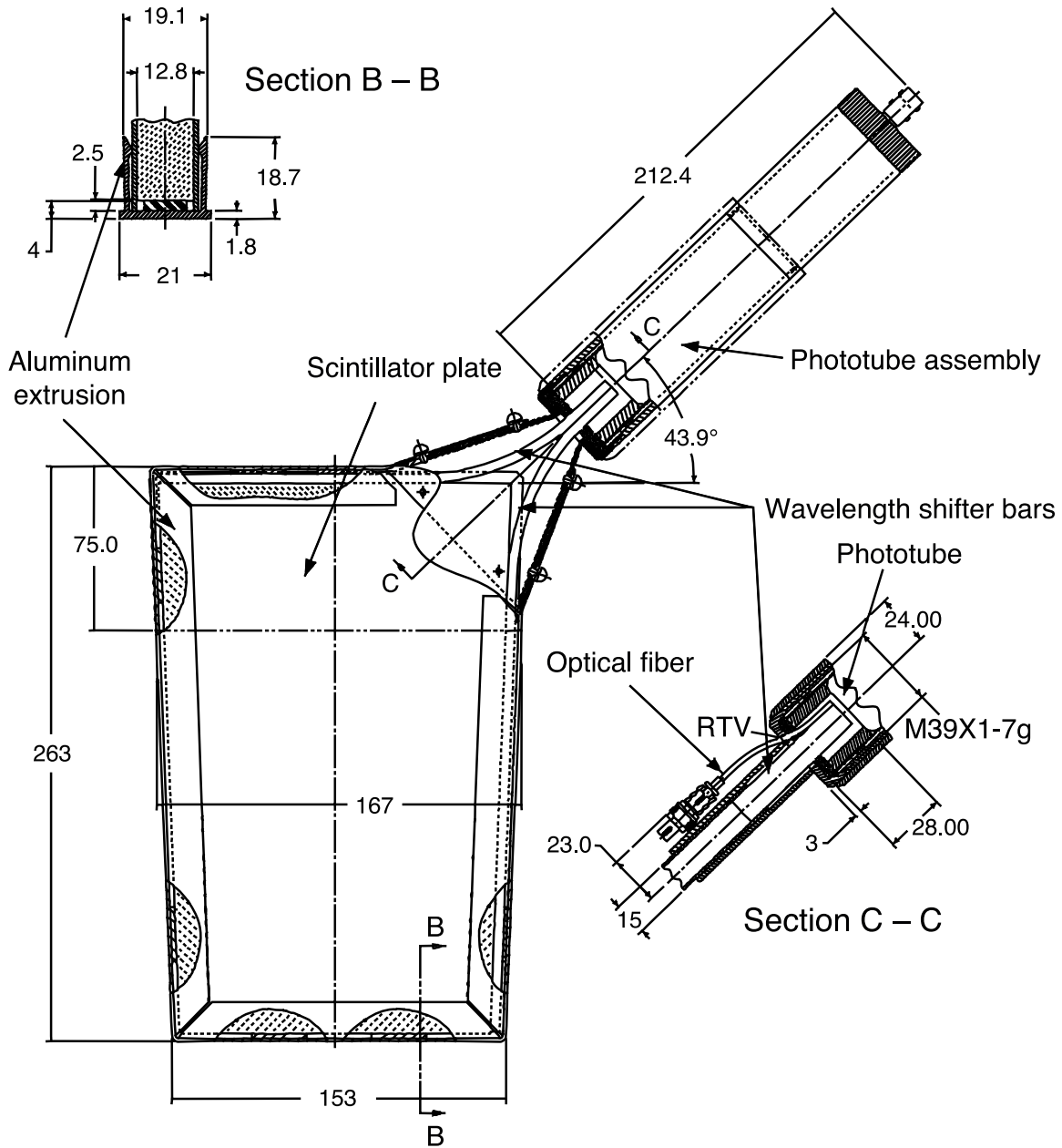}
\end{center}
\vskip 0.2cm
\caption{The design of a pixel scintillation counter. Dimensions are given in mm.}
\label{picpixdes}
\end{figure}

Fig.~\ref{picpixdes} shows the details of the counter design.  All four sides of the trapezoidal  
scintillator plate are machined but not polished.  Our studies concluded that the less expensive 
unpolished design has better photoelectron yield and uniformity \cite{refevdoki,refbez}. 
Two WLS bars, 4.2~mm thick 
and 12.7~mm wide, are placed along two edges of the scintillator plate with air gaps provided by 
narrow strips of adhesive tape attached to the scintillator plate as spacers. The end sections 
of both bars are bent by 44$^{\circ}$ to deliver light to the 25-mm-diameter MELZ 115M 
phototube \cite{refjoint}.  
The opposite ends of bars are made reflective using aluminized Mylar tape. To provide light 
tightness, the scintillator and WLS bars are wrapped in a layer of TYVEK material \cite{refdupont}
(type 1056D) and two layers of black paper. The wrapped counter assembly is secured in a box 
made of two aluminium plates and an extruded aluminium profile along the perimeter.  

A different light collection design was used for approximately 150 counters
located in cramped areas. These counters have twelve WLS fibers along the edges of 12.7-mm-thick 
Bicron 404A scintillator plates for collecting the scintillation light. These fibers are BCF 92 
multiclad fibers \cite{refbicron}. Both ends of the fibers are glued into plexiglass 
light-collector tubes, polished, and brought to a MELZ 115M phototube located on top 
of the counter assembly. This design provides 60 to 110 photoelectrons per minimum 
ionizing particle, depending on the counter's size. Using multiclad instead of singleclad 
BCF 92 fiber \cite{refevdoki} increases the light yield 1.4 and 1.7 times 
for counter 
sizes 13$\times$18~cm$^{2}$ and 38$\times$44~cm$^{2}$, respectively. 

A phototube assembly; containing a phototube, magnetic shielding, and tube base; is attached to the counter 
case by a threaded connection. The residual magnetic field of the muon toroids and the central solenoidal 
magnet is typically 150 G for A-layer pixel counters but can reach 300 G in some locations. 
For the B-layer pixel counters, the typical value is 100 G. The phototube magnetic shielding is similar 
to the shielding used for the phototubes of the A$\varphi$ counters. A resistive phototube base 
optimized for high gain and high time resolution is mounted at the end of the phototube.

\begin{figure}
\begin{center}
\includegraphics[width=1.0\textwidth,height=0.555\textwidth]{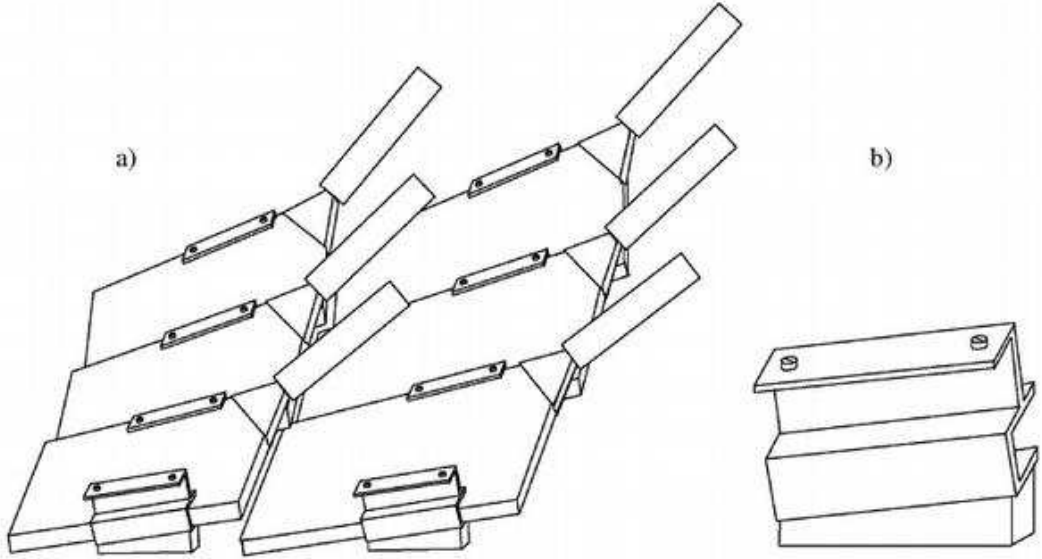}
\end{center}
\vskip 0.2cm
\caption{Fish scale design of the pixel counter mounting: a) counters arrangement; b) an aluminium mounting bracket.}
\label{picfish}
\end{figure}

The counters are installed on the frames using a ``fish scale'' design as shown in Fig.~\ref{picfish}. 
On the flat support frame, counters are tilted and supported by aluminium mounting brackets to achieve 
gapless coverage. All four edges of a counter overlap with the edges of neighboring counters. 
This novel design makes the procedure for assembling the pixel counter octants straightforward and 
also made access to phototubes easy. Eight octants are assembled into a pixel counter layer in 
the D$\O$ assembly hall and then a complete pixel counter layer is mounted on the detector as shown 
in Fig.~\ref{picpixphoto}. 
Placement accuracy of the octants is 3~mm and location of each layer is surveyed using optical methods
to 1~mm accuracy. After parts of the D$\O$ detector are moved to provide access for service
or repairs, the location of moved pixel counter layers is resurveyed with 1~mm precision. 

\subsubsection{Pixel counter performance}

\begin{figure}
\begin{center}
\includegraphics[width=1.0\textwidth]{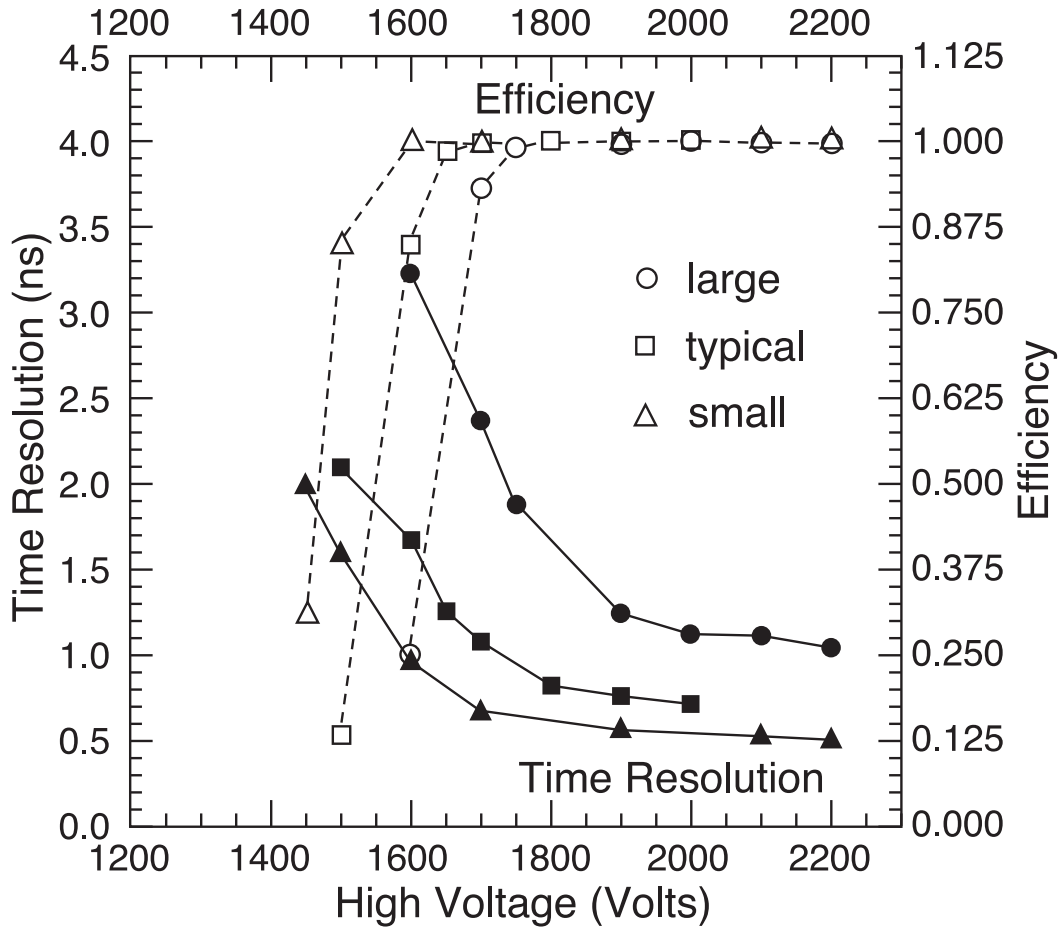}
\end{center}
\caption{Efficiency and time resolution of pixel counters. Open signs represent
efficiency, closed signs time resolution.}
\label{picpixeff}
\end{figure}

Performance studies have been done at the Fermilab 125~GeV/$c$ test beam using prototype counters 
with WLS bar readout.  Three different sizes; large (60$\times$106~cm$^{2}$), 
typical (24$\times$34~cm$^{2}$), and small (17$\times$24~cm$^{2}$); were tested. The measured 
efficiencies and time resolutions shown in Fig.~\ref{picpixeff} indicate that the pixel counters' peak 
efficiency is greater than 99.9\% and the time resolution, depending on counter size, 
ranges from 0.5~ns to 1~ns. The average light yield for 125~GeV/$c$ pions was measured to be 60 
photoelectrons for the large counter and 184 for the small counter.

\begin{figure}
\begin{center}
\includegraphics[width=1.0\textwidth,height=1.0\textwidth]{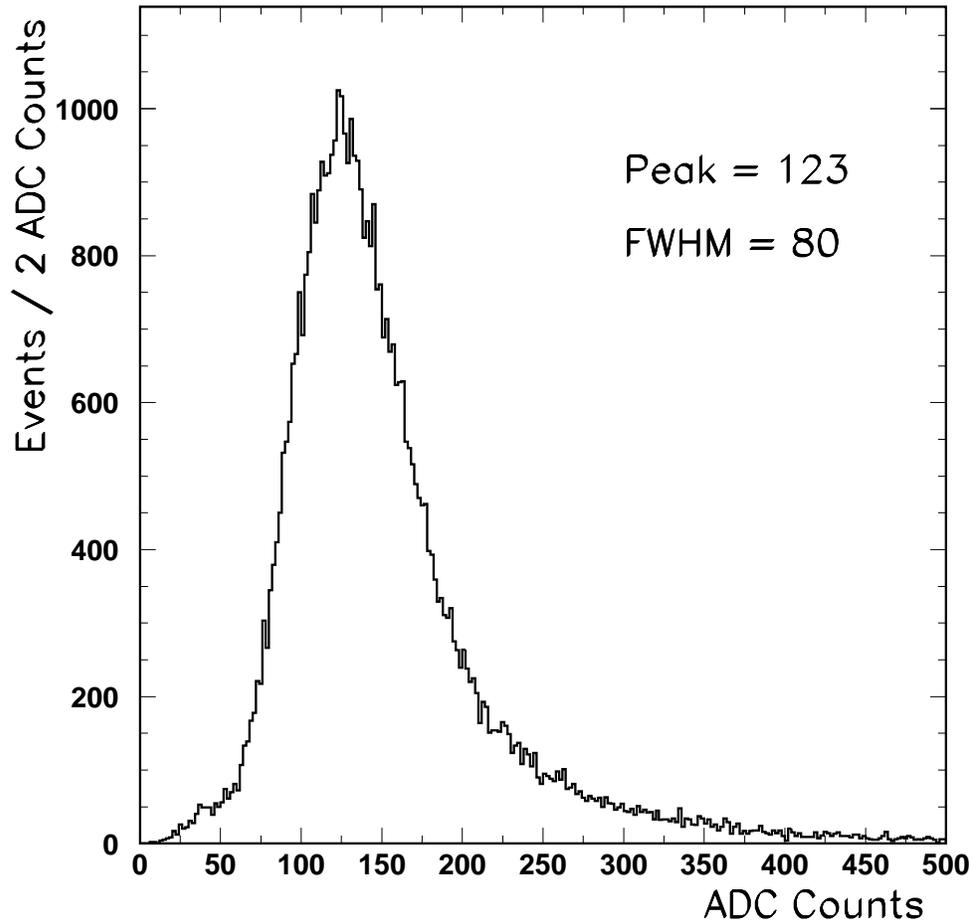}
\end{center}
\caption{Muon pulse height distribution for all 4214 pixel counters.}
\label{picpixamp}
\end{figure}

The production procedure for pixel counters included tests to check size tolerances and light yield 
measurements using a $^{60}$Co radioactive source. Some counters of each size have been tested 
using cosmic ray muons to calibrate radioactive source measurements to the number of photoelectrons. 
According to the cosmic test, the number of photoelectrons varies from 240 for small A-layer 
counters to 70 for the largest C-layer counters. The uniformity of light collection for particles
irradiating the full surface of the counters is $\pm$10\%. Phototube-to-counter matching has been performed 
to get close signal amplitudes for each counter connected to the same high voltage power supply 
in a group of sixteen. This matching included the use of phototube test stand results obtained using 
LED and $^{60}$Co measurements. A final matching check was made by pulse height 
measurements using a $^{90}$Sr beta source. The uniformity of $^{90}$Sr amplitudes for groups of sixteen counters
achieved using the matching procedure is $\pm$25\% (full width). 
High voltage values selected during the matching process are used for Run~II data taking. 
Uniformity in signal amplitudes 
as well as a reasonable choice of discriminator thresholds are illustrated in Fig.~\ref{picpixamp}. 
This figure shows the amplitude 
distribution for reconstructed muons from proton-antiproton collisions for all 4214 counters.
The discriminator threshold is 
equal to 7~mV or about 25 ADC channels.

\begin{figure}
\begin{center}
\begin{flushleft}
\hskip -1.0in
\begin{tabular}{p{160mm}}
\includegraphics[width=1.4\textwidth]{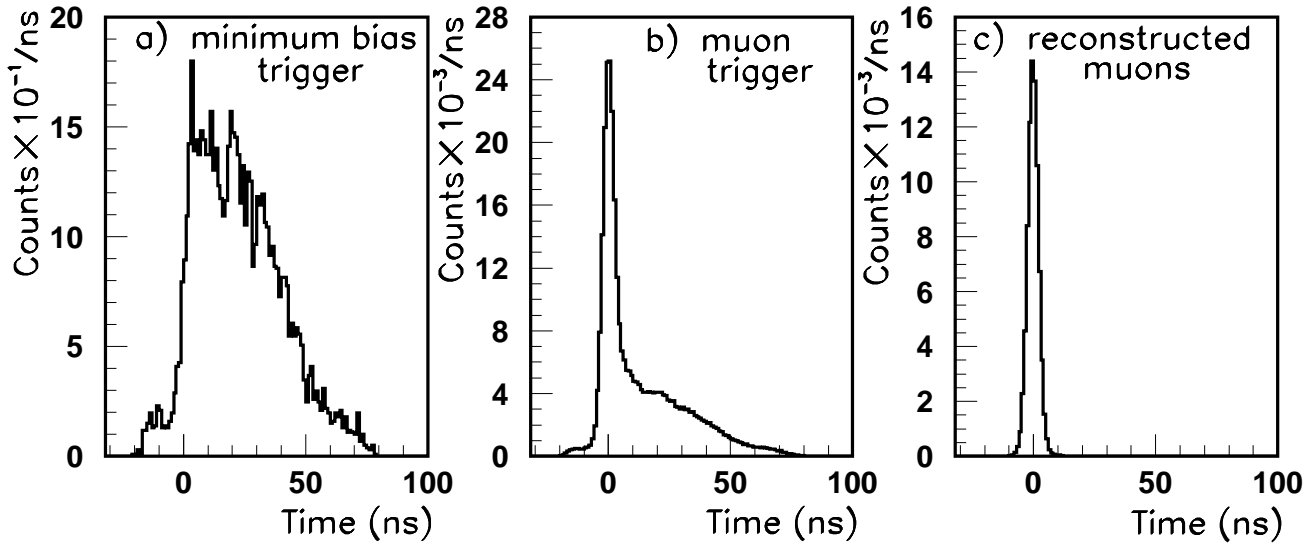}
\end{tabular}
\end{flushleft}
\end{center}
\vskip -0.3cm
\caption{Time spectra of hits recorded by pixel counters: a) minimum bias trigger; b) muon trigger; c) reconstructed muons.} 
\label{picpixtime}
\end{figure}

Time distributions of pixel counters hits during proton-antiproton collisions run are 
shown in Fig.~\ref{picpixtime}. The time 
distribution of muon hits originating in the collision region peaks at zero. The time 
distribution of events selected with a minimum bias trigger, Fig.~\ref{picpixtime}a,
has a broad peak that ranges from zero to about 70 ns, mainly due to the late-arriving background 
particles discussed in Section 3. 
Hits in the 30~ns time gate are sent to the Level 1 trigger.
The distribution obtained for events selected 
with a Level 1 muon trigger, Fig.~\ref{picpixtime}b, shows a peak at zero with some background remaining. 
The time distribution for hits 
from reconstructed muon tracks, Fig.~\ref{picpixtime}c, has a clean peak at zero with a time 
resolution of all counters combined  
\begin{math}
\sigma_{t}=2.2 
\end{math}
ns. These results demonstrate the 
power of the D$\O$ forward muon trigger counters for selecting muons in high rate 
proton-antiproton collisions. In D$\O$, about 50\% of Level 1 forward muon triggered 
events have a muon reconstructed off-line. Additional rejection from the Level 2 trigger provides the 
capability of recording forward muon data sample with 80\% purity and efficiency of about 90\%.  

\begin{figure}
\begin{center}
\includegraphics[width=1.0\textwidth]{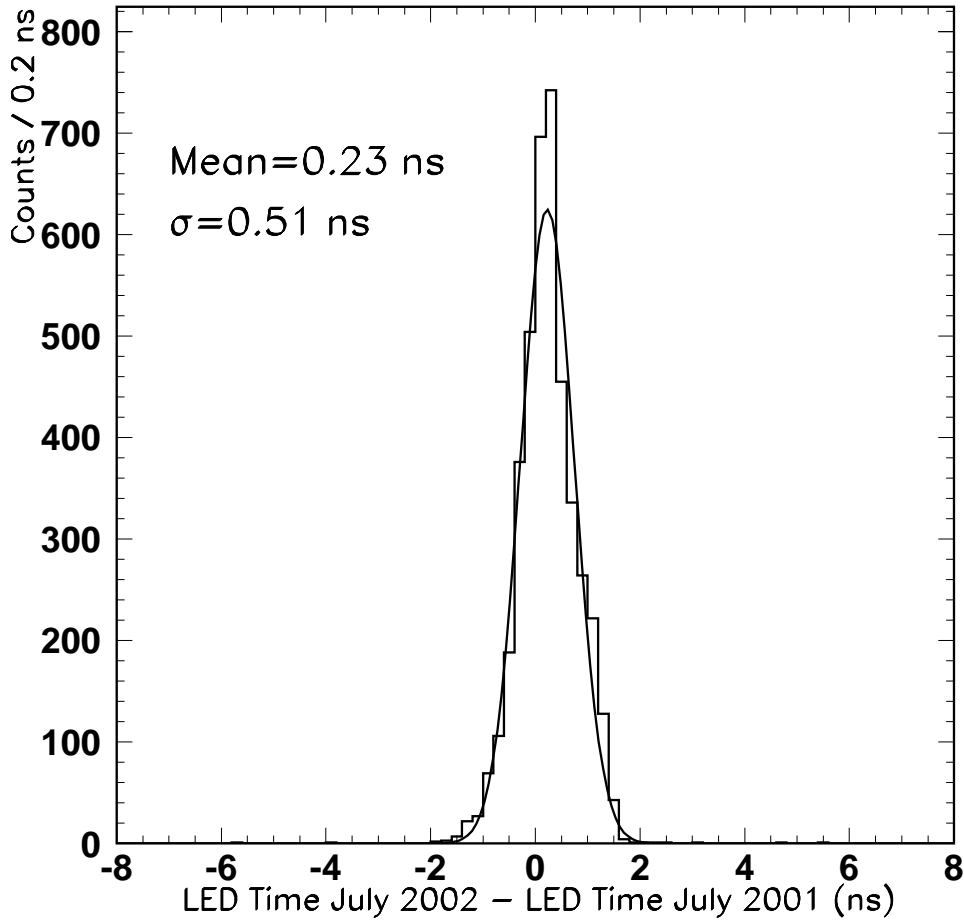}
\end{center}
\caption{Timing stability of pixel counters monitored by the LED calibration system over one year.}
\label{picpixled}
\end{figure}

Pixel counters are demonstrating stable operation and high reliability during physics data collection.
Figure~\ref{picpixled} shows the change in timing of 4214 counters in a one year interval measured by the 
LED calibration system (see Section 4.5). Timing is stable within 0.5~ns for all 4214 counters.  The stability of the counter amplitudes is also measured by the LED pulser system. 
We found that amplitude stability is about 10\% for individual counters in three years
of operation. 
The stability of the signal amplitude, averaged over all phototubes, is better than 1\%. 

About fifteen counters have failed over three years of operation primarily due to bad 
connectors and bases. 
Long-enough access to the collision hall to repair failed counters occurs every 3$-$6 months.
Movers designed to roll out fully assembled A and B-layers of the pixel counters to provide
access to all detector parts for repairs are mounted on the top of the layers. The typical duration of 
opening/closing a pixel counter layer including repairs is one day.
The typical number of non-working counters during physics data taking is about two or 0.05\%.

\subsection{High voltage and readout electronics for scintillation counters}

A total of 46 negative high voltage power supply modules \cite{refhv} located in a VME crate provide high voltage 
to the 630 A{\it $\varphi$} counters in the central region. Another 288 negative high voltage power 
supply modules placed in six VME crates supply high voltage to the 4214 forward pixel counters. 
The power supplies are rated at 3.5~kV with a maximum current 3.0~mA. Typical current per phototube 
base is 145 $\mu$A at 1.85~kV and the total current per power supply is approximately 2~mA. 

Signals from phototubes are transmitted via 50 $\Omega$ coaxial cables to the VME based front-end 
electronics. This electronics system is discussed in more detail 
in Section 6. The amplified signals are sent to a 10-bit ADC (one ADC for 16 channels, multiplexed) 
and to discriminators with a variable threshold (1$-$250 mV range, individually adjustable for each channel). 
Discriminated and gated signals from the fired counters are sent to the Level 1 trigger electronics 
system and to the TDCs with a 1.03~ns bin size. After digitization, the amplitude and time information 
is sent to the Level 2 trigger system and the D$\O$ data acquisition system. 

\subsection{Scintillation counters monitoring and calibration}

Calibration of various electronic components is performed periodically by using precision pulsers 
that inject charge directly into electronics inputs. In addition to the electronics calibration, a specially designed 
LED-based calibration system was developed for monitoring the timing and gain of counters \cite{refhanlet}. 
This system illuminates phototubes with light pulses similar to those of muon signals in amplitude and shape. 
The light pulses are generated by two types of LED pulser modules. 
Blue LED pulser modules distribute light signals to 100 separate fiber optics cables 
which are connected to MELZ phototubes. 
Modules using blue-green LEDs have 54 fibers to distribute the light pulses
which are connected to EMI phototubes. 
 
\begin{figure}
\begin{center}
\begin{flushleft}
\includegraphics[width=2.1\textwidth]{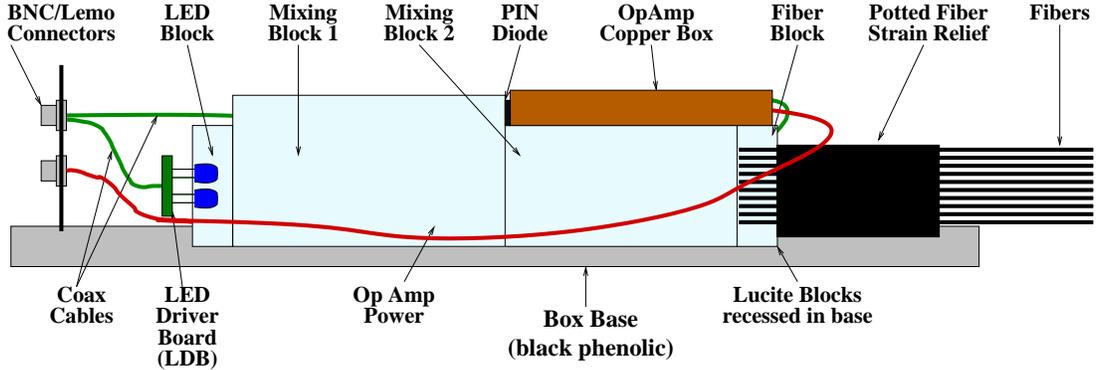}
\end{flushleft}
\vskip -4cm
\end{center}
\caption{Design of the LED pulser module.}
\label{picleddes}
\end{figure}

A sketch of an LED pulser module is shown in Fig.~\ref{picleddes}. An external pulser sends 
signals to the LED driver board that generates current pulses to drive the LED. 
The waveform of the current pulse creates a signal that imitates a muon
passing through the counter. 
To create a uniformly illuminated field, the light pulse generated by the LED goes through 
two stages of light mixing blocks. 
A PIN diode is
mounted on the first light mixing block to provide an internal measure of the light intensity being
distributed to the phototubes. The light pulses are split by the fiber block and fed to phototubes via clear plastic optical fibers
with 1~mm diameter. The light splitter is located directly downstream of the second mixing block 
as shown in Fig.~\ref{picleddes}. An optical connector is mounted on each counter for connecting the optical fiber.

The performance of the scintillation counters is monitored
on a regular basis during collider runs. Voltages and currents of the phototube high voltage power 
supplies are checked during every data taking shift using the D$\O$ online monitoring system. 
Using large samples of muons produced in the proton-antiproton collisions, hit distributions, 
efficiencies, signal amplitude and timing distributions of counters are monitored online. 

\section{Muon tracking system}

The Run~II D$\O$ muon tracking system consists of the original PDT chambers in the central region and 
new MDT detectors in the forward region. The muon momentum resolution in the central rapidity region 
is dominated by the inner tracking system which includes the silicon microstrip tracker and the 
scintillating fiber tracker. 
The muon tracking system provides muon identification and an independent, confirming 
momentum measurement. The MDT system in the forward region plays a more important role in determining 
the muon momentum for $|$$\eta$$|$ $>$ 1.6 where the performance of the central tracker is reduced.

\subsection{Central muon tracking system}

The central muon tracking system consists of 94 PDT chambers with 6624 cells 
covering the region $|$$\eta$$|$ $<$ 1.0. As shown in Fig.~\ref{picd0}, the PDT chambers are grouped into 
three layers. The inner A-layer chambers are mounted on the inside surfaces of the central iron toroidal 
magnet. The B-layer and the C-layer chambers are placed outside the toroidal magnets. 

The PDT chambers are assembled from three or four decks of proportional drift tubes \cite{refbrown}. Each proportional 
drift tube consists of a rectangular shaped aluminium enclosure, an anode wire at the center and two 
cathode pads above and below the anode wire. The cathode pads are made of thin copper-clad 
Glasteel \cite{refglasteel} strips with etched copper vernier pads to determine the longitudinal position 
of the muon hits. The copper-clad Glasteel is made of glass-fiber reinforced polyester with catalysts added 
to promote copper bonding. 
The Glasteel electrodes outgas and, in a high radiation environment, coat the anode wire
with a sheath of insulating material. A procedure for removing the sheath material \cite{refmarsh}
was developed in Run~I that returns the chambers to nominal performance. Because access to the A-layer 
PDTs is difficult, we replaced the electrodes in those chambers with copper-clad G-10 as part of the
Run~II upgrade. 

A fast, non-flammable gas mixture consisting of 84\% Argon, 8\% CF$_4$, and 8\% CH$_4$ is used for the PDT 
chambers in Run~II. The maximum electron drift time in the 10~cm wide drift cell is 450~ns. With this 
gas mixture, anode wires are operated at 4.7~kV and the cathode pad electrodes at 2.3~kV. The single-wire resolution 
is approximately 1~mm due to the electron diffusion in the drift cell with its 5~cm half-cell 
width and signal amplitude fluctuations.

The total gas volume of the PDT chambers is 10$^2$~m$^{3}$ and the gas mixture is recirculated. 
The design of the gas circulation system takes into account the outgasing problem due 
to the Glasteel electrodes in the B and C-layer PDTs by providing two separate gas circulation systems, 
one for the A-layer PDT chambers and the other for the B and C-layer PDT chambers. 
The recirculated gas mixture in both systems goes through molecular filters that removes 
contaminants from the Glasteel electrodes and other sources. Both gas circulation systems are able to 
circulate the gas mixture rapidly to remove the contaminants. The average 
flow rate per PDT chamber is 2 liters per minute resulting in one volume change in about 8 hours.
The wire cleaning procedure has not been necessary after accumulating 0.8~fb$^{-1}$ luminosity 
during Run~II due to increased gas flow, improved shielding, and installation of G-10 electrodes
in A-layer chambers.

\subsection{Forward muon tracking system}

During the Run~II upgrade, the original PDT chambers in the forward region 
(1.0 $<$ $|$$\eta$$|$ $<$ 2.0) were replaced by a new forward muon tracking system. 
Increased luminosity and reduced beam bunch spacing are the two main reasons for building this new system. 
The new D$\O$ forward muon tracking system is based on a mini drift tube technology similar to the one 
described in Ref.~\cite{refbusza}. 

\subsubsection{Mini drift tubes}

\begin{figure}
\begin{center}
\includegraphics[width=1.1\textwidth]{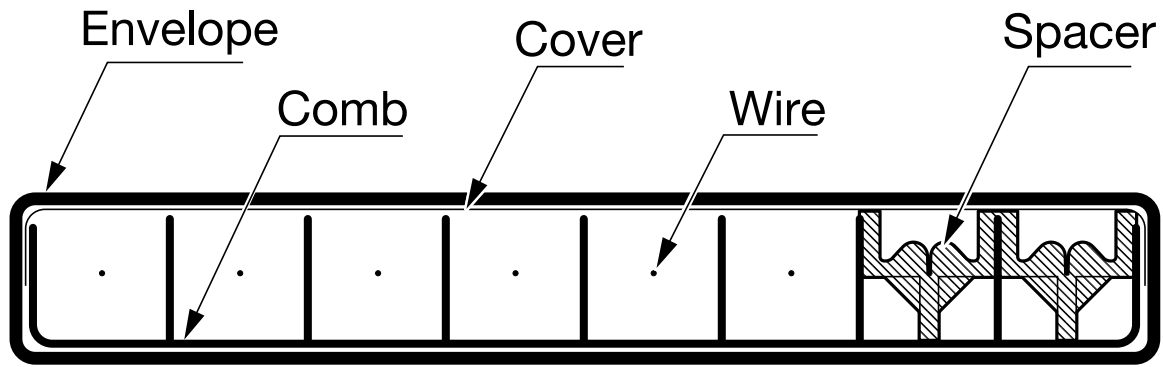}
\end{center}
\caption{The cross-sectional view of a mini drift tube.}
\label{picmdt}
\end{figure}

\begin{figure}
\includegraphics[width=1.0\textwidth]{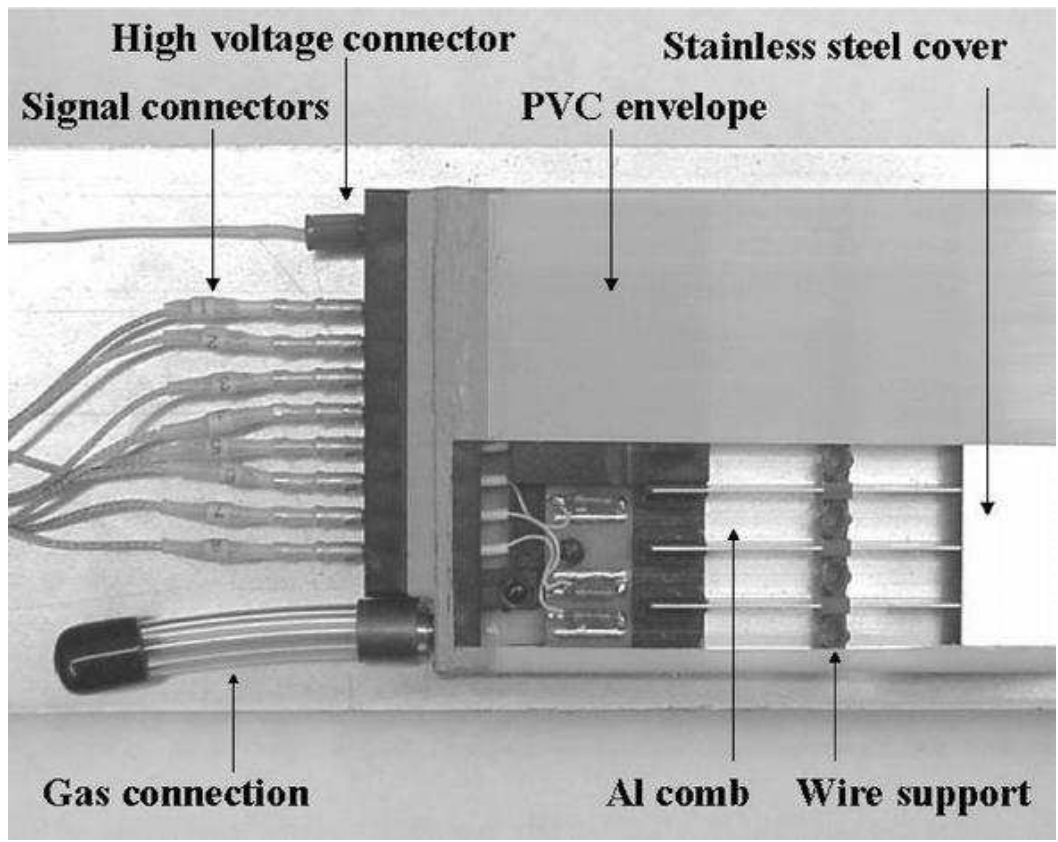}
\caption{The signal and high voltage end of a mini drift tube with cover partially removed.}
\label{picmdtphoto}
\end{figure}

A cross-sectional view of a mini drift tube is shown in Fig.~\ref{picmdt}. Each mini drift tube has 
eight 50-$\mu$m gold-plated tungsten
anode wires suspended in a comb-shaped thin-wall aluminium profile. The wire tension is 200~g and 
the wire spacing is 10~mm. 
The wires are supported by plastic spacers at intervals of one meter.
The wall thickness of the aluminium profile is 0.6~mm. A 0.15-mm-thick stainless steel cover, 
placed on top of the aluminium profile, completes
the cathode enclosure for the eight 9.4$\times$9.4~mm$^{2}$ drift cells. 
The metal structure is enclosed in a 1-mm-thick plastic envelope 
made of polyvinyl chloride (PVC) that provides the electrical insulation for the metal 
cathode operated at negative high voltage. Two plastic
end plugs are bonded to the two ends of the PVC envelope forming a gas-tight enclosure. Fig.~\ref{picmdtphoto} 
shows a mini drift tube with its cover partially removed. The lengths of the tubes range from approximately 1~m to 6~m. 
The mini drift tubes are operated in the proportional mode and drift times from the anode wires are recorded.

The gas mixture for the MDT system is 90\% CF$_{4}$ and 10\% CH$_{4}$. The selection 
of this mixture as the working gas for the MDTs was based on 
the following considerations. The gas mixture must be non-flammable for safety reasons. The electron 
drift speed must be fast enough so that the maximum drift time plus the signal propagation 
time in the longest drift tube is less than the beam bunch spacing. And, the gas mixture must not 
result in wire aging in a high radiation environment. 
The 90\% CF$_{4}$$/$10\% CH$_{4}$ mixture satisfies all these requirements. The muon efficiency 
plateau is measured to be approximately 0.40~kV (from $-$3.0~kV to $-$3.4~kV). The operating MDT  
voltage is $-$3.2~kV with a signal discrimination threshold of 2.0~$\mu$A and a muon detection 
efficiency greater than 99\%. 

Aging tests using radioactive sources have shown no detectable aging effects up 
to 2.5~C/cm accumulated charge on the anode wire. The total accumulated charge 
for an integrated luminosity of 10~fb$^{-1}$ is 
expected to be 0.02~C/cm in the MDTs closest to the beam line. 

\begin{figure}
\begin{center}
\includegraphics[width=1.0\textwidth]{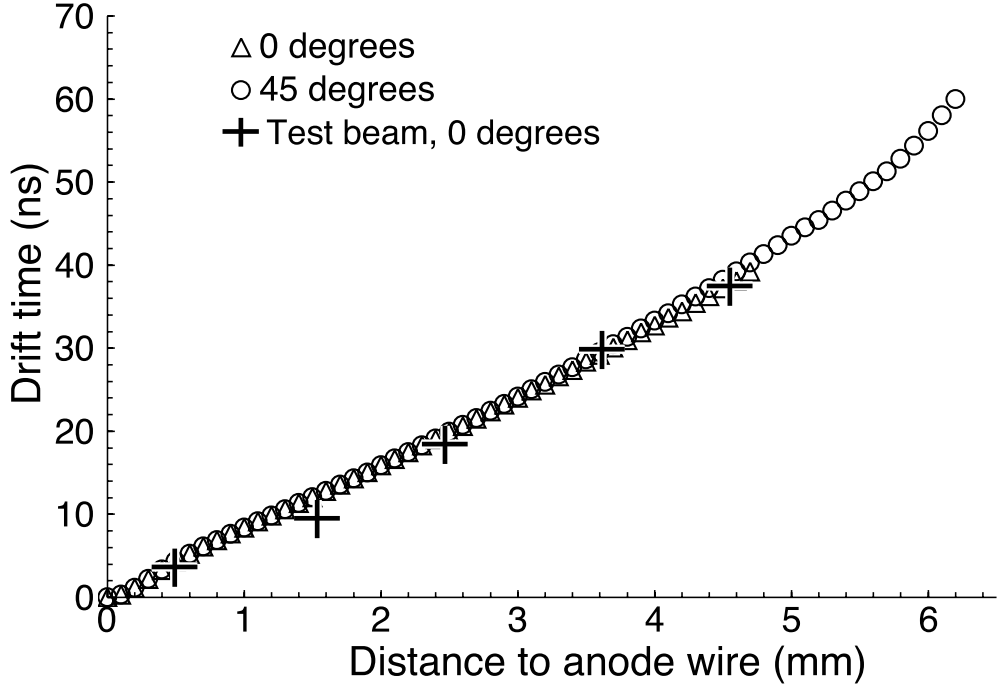}
\end{center}
\caption{Time-to-distance relation for a mini drift tube. The data labeled as triangles and circles 
are calculated using GARFIELD. The gas mixture is 90\% CF$_{4}$ and 10\% CH$_{4}$ 
and the cathode voltage is $-$3.2 kV.} 
\label{picmdttime}
\end{figure}

The electron drift time-to-distance relations calculated by the simulation 
program GARFIELD \cite{refgarfield} and experimentally measured are shown in Fig.~\ref{picmdttime}. 
To investigate the effect of the square cell geometry, the time-to-distance relations were calculated 
for two drift paths in a drift cell of the MDT. The data labeled with triangles is for tracks that enter the drift cell perpendicular to the surface of the MDT detector plane. 
The data labeled with circles is for tracks entering along the diagonal direction of the 
square cell. The maximum drift distance for 0$^{\circ}$ 
tracks is 4.7~mm and the calculated maximum drift time is approximately 38~ns. In the 45$^{\circ}$ case, 
the maximum drift time for a drift distance of 6.2~mm is about 60~ns. As shown in Fig.~\ref{picmdttime}, 
the 0$^{\circ}$  and 45$^{\circ}$ curves mostly coincide. This implies that different drift fields in these 
two directions do not significantly affect the electron drift velocity. 

\subsubsection{MDT system design and construction}

The MDT system consists of 6080 mini drift tubes that are assembled into six layers of eight octants each. 
The total number of anode wires is 48,640. Vigorous quality assurance procedures were implemented 
at every level of the construction. Before the MDTs were assembled into octants,
the anode wire tension, the response uniformity, and the gas tightness were tested. Cosmic ray tests 
were performed for every MDT to verify its operational characteristics.
MDTs that passed quality tests were mounted on frames made of aluminium 
honeycomb plates and then assembled into octants. The flatness of the aluminium honeycomb plates was 
required to be within 1~mm over their entire surfaces. The MDTs are held in place by
rows of plastic blades. The spacing of the blades is tightly controlled to ensure the 
tube's placement accuracy. The outside surfaces of an octant are enclosed by thin aluminium sheets that 
form a Faraday cage and also protect the MDTs mechanically. Gas tightness and cosmic 
ray tests were performed for every completed octant.

The forward muon tracking system is arranged as A, B and C-layers similar to the configuration of the 
pixel counter system discussed in Section 4.3.1. Each layer has eight octants. An A-layer octant contains 
four planes of tubes with each plane having 32 MDTs. The four tube planes 
are mounted on two aluminium honeycomb frames. Each B-layer octant has
three MDT planes and each plane consists of 44 MDTs, except the two bottom 
octants that have only 30 MDTs. Each C-layer octant has three planes 
and each plane consists of 48 MDTs, except the two bottom octants that 
have only 30. The parameters of the MDT system are summarized in Table~3. 

\begin{table}
\begin{center}
Table 3 \\
Channel count for MDT system. \\
\vskip 2mm
\begin{tabular}{|l|c|c|c|} \hline
Layer&A&B&C \\ \hline
Number of MDTs&2048&1944&2088 \\
Number of readout channels&16384&15552&16704 \\
$\eta$$_{\rm{max}}$&2.15&2.13&2.16 \\
$\eta$$_{\rm{min}}$&1.00&1.02&1.13 \\
Maximum tube length (mm)&3571&5066&5830 \\ \hline
\end{tabular}
\end{center}
\end{table}

A-layer MDT octants are mounted on the inside faces of the end iron toroids and B-layer octants are 
mounted on the outside faces. The C-layer octants are mounted on two steel structures placed against 
the walls of the collision hall (Fig.~\ref{picd0}). 
Placement accuracy of MDT octants on the detector is 5~mm. After octants are mounted in the detector,
their locations are measured using an optical survey with 0.5~mm precision. After parts of the D$\O$
detector are moved to provide access for service or repairs, the position of MDT octants is re-surveyed. 
The MDTs in each layer are oriented in such a way 
that the bending angle in the toroid magnet of a muon's trajectory can be measured precisely. 
A crude determination of the muon trajectory in the direction transverse to the magnetic bend 
direction is provided by the pixel scintillation counters.

Amplifier-discriminator boards (ADBs) \cite{refalexeev} are mounted near the edge of the octant closest to the 
MDT signal end plug on the aluminium covers. The ADBs are protected and shielded by aluminium 
enclosures. Each ADB has 32 channels that amplify and discriminate the signals from the 32 anode wires. 
The output differential signals of the ADBs are sent via flat cables to digitization boards 
in VME crates. 

\subsubsection{Gas system}

The MDT gas system consists of the gas recirculation, flow rate monitoring, and gas 
purification systems \cite{refzhao}. The total gas volume of the MDT system is 18~m$^{3}$. 
The CF$_{4}$/CH$_{4}$ mixture is recirculated to minimize operating cost 
with a flow rate of 6~{\it l}/min. This flow rate provides a gas 
exchange rate of approximately 0.5 volumes per day. The gas output pressure of the octants is kept 
to approximately 5~cm of H$_{2}$O above atmospheric pressure. A gas purification 
system that removes water vapor and oxygen in the return gas mixture prevents the 
accumulation of such impurities during gas circulation. Fresh gas is added into the MDT 
gas circulation at an average rate of 0.5~{\it l}/min to compensate for losses due to small leaks 
in the system and to refresh the gas.

The gas flow through the MDTs in each plane (up to 48 tubes) is in series. Gas flow 
sensors are installed at the gas inlet and outlet of each MDT plane within an octant 
to monitor the flow rates. There are 160 planes of MDTs (four in each A-layer octant
and three in each B and C-layer octant) and 320 gas flow sensors are 
used. The sensors are model AWM3300V mass flow sensors made by Honeywell Inc. \cite{refzhao}. 
These flow sensors are mounted on the octants. The flow rates of the 48 MDT octants 
can be individually adjusted by precision valves.

\subsubsection{MDT system performance}

The MDT muon tracking efficiency measured during collider operation is greater than 99\% in the active gas volume of the tube cells. 
The average per plane muon hit detection efficiency is (95 $\pm$ 2)\%.
Inefficiency of the MDT planes is mainly caused by geometric factors that include the
0.6~mm thick aluminium comb wall between drift cells and gaps between adjacent tubes 
(1~mm thick plastic blade plus the walls of the two tubes). 
At the beginning of operation in 2001, 0.16\% of the MDT system channels were non-functional. 
After three years of running, the percentage of dead channels has increased to approximately 0.24\%.  
The major sources of non-working channels are broken and noisy wires.

The coordinate accuracy of the MDT system is limited by the 18.8~ns bin size of the 
drift time digitization. The design goal of the single cell resolution of the MDT system is 1~mm. 
The single cell resolution can be measured by fitting a straight line through the three muon 
hits detected in the three or four planes of an MDT octant. 
Coordinate resolution obtained using this method is $\sigma$ = 0.8~mm.

\section{Muon front-end and readout electronics}

Design of the Run~II front-end and readout muon electronics is based on the following requirements:

\begin{itemize}
\item the ability to run with a 132~ns bunch spacing (original specification for Run~II, later changed to 396~ns)
\item continuous digitization of the input signals 
\item digital pipelines to store data during the Level 1 trigger decision time 
\item deadtimeless Level 1 trigger processing
\item use of DSPs for data preprocessing and data buffer management 
\item high speed serial cable links to transfer trigger and data information
\item stand-alone calibration and test capabilities.
\end{itemize}

\begin{figure}
\begin{center}
\begin{flushleft}
\begin{tabular}{p{160mm}}
\includegraphics[width=1.0\textwidth]{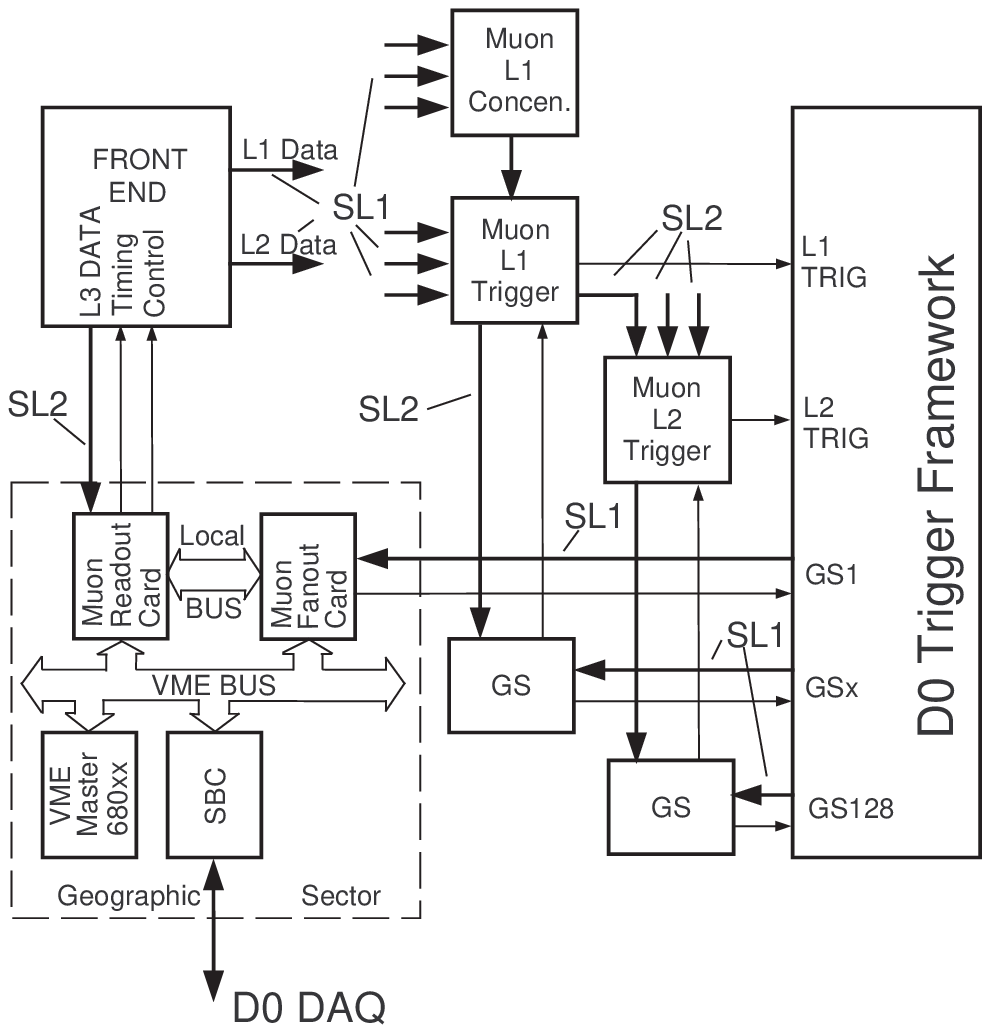}
\end{tabular}
\end{flushleft}
\end{center}
\caption{Block diagram of the D$\O$ muon readout electronics.}
\label{picreadout}
\end{figure}

All muon detector subsystems have a common readout interface based on a 
commercial VME processor and two custom VME modules: the muon readout card (MRC) and the muon 
fanout card (MFC) described in Ref.~\cite{refbaldin}. The readout electronics is housed in 
eighteen 9U$\times$280~mm VME crates. A block diagram of the muon electronics is shown in Fig.~\ref{picreadout}. 

The D$\O$ trigger framework (TFW) is the source of timing and control signals for the 
entire experiment. The detector electronics is subdivided into Geographic Sectors (GS) associated 
with the front-end detector electronics. Timing and trigger 
information is distributed to the GSs using 1~Gbit/s serial links (SL1). 
Each link is capable of transferring 96 bits of information within 132~ns. 
A similar 1~Gbit/s link is used to transfer information from the front-end electronics to the 
Level 1 trigger system. 

All of the front-end crates are synchronized to the accelerator RF frequency
(53.1047~MHz) allowing for accurate timing of the electronics to the collider beam. 
Detector signals are continuously digitized by charge-integrator/ADC or 
discriminator/time-digitizer signal processing chains and stored in circular buffers (pipelines) 
thus achieving deadtimeless operation \cite{refbaldin}. The depth of the buffers is adjusted to match 
the propagation time of the Level 1 trigger decision. Upon arrival of the Level 1 trigger decision, 
the stored information is copied to the Level 1 FIFO and awaits the next level of the trigger decision. 
Depending on the Level 2 trigger decision generated by the TFW, the stored information is processed 
and transferred to the MRC using a 160~Mbit/s serial link (SL2) or discarded. From the MRC memory, 
event data is sent to the D$\O$ data acquisition system under control of the VME processor to be
analyzed by the Level 3 trigger system before selected events are written to tape.

\begin{figure}
\begin{center}
\includegraphics[width=1.0\textwidth]{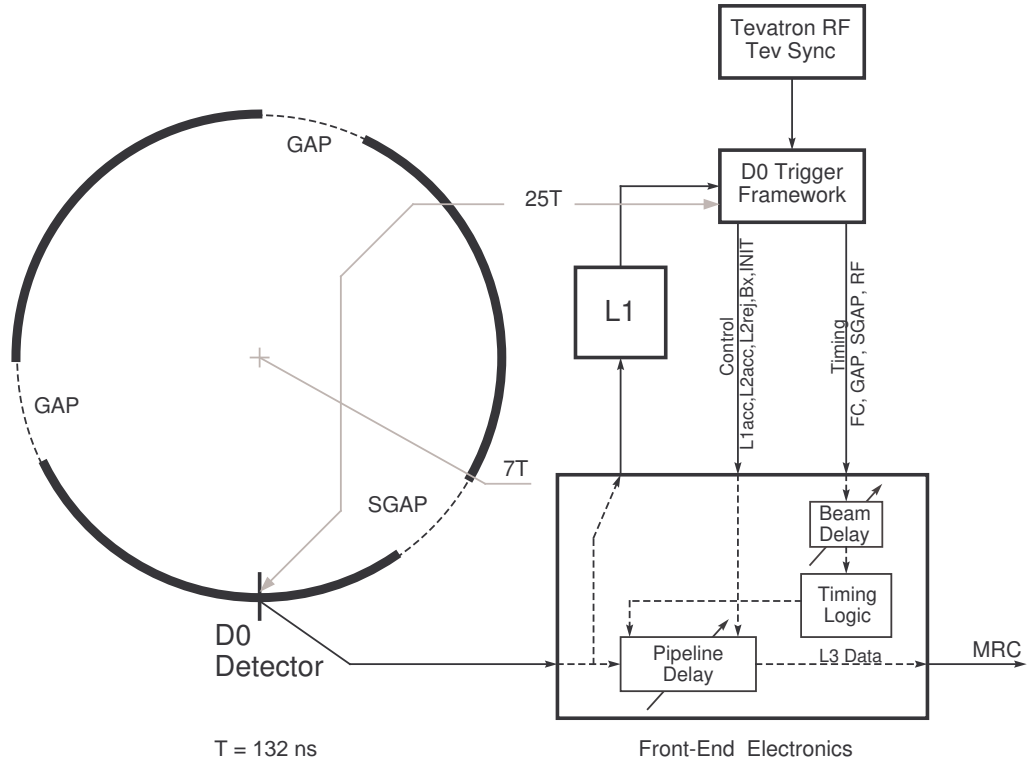}
\end{center}
\caption{D$\O$ muon detector timing. The circle represents the Tevatron ring filled with particle
bunches (solid line) with abort and synchronization gaps.}
\label{pictiming}
\end{figure}

Collider beam bunches are separated by 21 periods of the RF 
frequency (396~ns). Seven periods of the RF frequency (132~ns) are used as a unit of TFW timing. 
There are 159 such intervals 
around one accelerator orbit starting from the beginning of the synchronization gap (Fig.~\ref{pictiming}). 
The D$\O$ experiment uses the Tevatron crossing number scheme 
to guarantee proper event synchronization. The seventh crossing after the beginning of synchronization gap
contains the first particle bunch.
The specified delay between a beam collision and the 
arrival of the trigger signal from the Level 1 system at the TFW input is twenty-five 132 ns intervals.

One of the important features of the muon electronics design is the achievement of proper synchronization of 
all the elements of the system. To do this, the muon electronics has two variable digital delays 
implemented in the front-ends. The D$\O$ TFW receives timing information from the accelerator 
and generates two sets of timing signals representing the timing structure of the colliding beams: 
one for its own logic and the other for the D$\O$ detector subsystems (see Fig.~\ref{pictiming}). The latter has 
a timing offset to compensate for propagation delay in the cables. The offset is chosen so 
that timing signals arrive at the front-end electronics at approximately the same time as the corresponding 
detector signals arrive at the electronics inputs. Fine adjustments have to
be done using the so-called beam delay. The pipeline delay setting is affected by the beam delay setting 
and vice versa. To overcome this difficulty, a procedure involving the internal test pulser is used 
to set the correct pipeline delay value. 
After the pipeline delay is selected, the beam delay value is adjusted using beam data to compensate for 
cable delay differences. Any change in the beam delay adjustment must be followed by an equal 
change of the pipeline delay.

The muon front-end electronics includes PDT electronics, scintillation counter electronics, and MDT 
electronics. The readout
controllers for these subsystems have similar designs and an Analog Devices ADSP-21csp01 50 MHz DSP. 
The time digitizers used in the PDT and scintillation counter electronics are based 
on Toshiba TMC-TEG3 chips \cite{refarai}. 

The PDTs have up to 96 signal wires per chamber. Attached to these chambers are four 24-channel front-end boards 
(FEB) and one control board (CB). The FEBs digitize the time of arrival of the wire signal and the charge 
of the pad electrodes. The TMC chip is running at RF/2 frequency, providing 1.18~ns bin resolution. 
The digital delays and Level 1 FIFO buffers are also located on FEBs. Each FEB is connected to 
the CB via an 18 bit uni-directional data bus and an 8-bit control bus. The CB has a readout controller, 
which fetches data sequentially from each FIFO. Upon receiving a Level 1 trigger, the controller accepts data 
and stores it in the DSP memory.

The front-end electronics of the scintillation counter system includes 48-channel 9U VME 
scintillation counter front-end cards (SFE).
These cards measure the arrival time of the scintillation counter signals. The TMC chip runs at 4/7 
RF frequency with a timing bin resolution of 1.03~ns. There are three charge integrators on each SFE with outputs 
digitized by ADCs. Each integrator input can be connected (multiplexed) to one of sixteen input channels 
to measure pulse height from the photomultiplier tube. The Level 1 FIFO buffers are 
read out by a 
DSP-based readout controller over a custom high speed data bus using the VME J2 backplane. Each VME 
crate houses up to ten SFE cards, the readout controller card, LED calibration electronics, and a 
Motorola 68EC040-based \cite{refmotorola} VME processor to provide parameters downloading and testing.

The front-end electronics cards (ADBs) of the MDTs are located on the MDT chambers near the 
signal ends. Each ADB has 32 channels of amplifiers and discriminators. The digitization of the MDT wire 
signals is accomplished by a VME-based system. The digitization cards are 9U VME cards. Each card 
contains 192 channels. The MDT digitizing cards (MDCs) perform a low resolution (18.8~ns/bin) 
measurement of the drift time. The MDC also has a digital pipeline and Level 1 FIFO buffers to store event data. 
A DSP-based readout controller reads out data over a custom VME J3 back plane. The MDT crates can 
accommodate up to twelve MDCs, a readout controller, and a VME processor to perform tasks similar 
to those of the scintillation counter system.

\begin{table}
\begin{center}
Table 4 \\
Channel counts of the D$\O$ Run~II muon system. \\
\vskip 2mm
\begin{tabular}{|l|c|c|c|c|} \hline
Item&Layer A&Layer B&Layer C&Total \\ \hline
PDT chambers&18&38&38&94 \\
Drift cells&1584&2424&2616&6624 \\ \hline
Central scintillation counters&630&96&276&1002 \\
PMTs&630&192&552&1374 \\ \hline
MDT octants&16&16&16&48 \\
Drift cells&16384&15552&16704&48640 \\ \hline
Pixel octants&16&16&16&48 \\
Counters&1518&1420&1276&4214 \\ \hline
\end{tabular}
\end{center}
\end{table}

\vskip 3.0cm

\section{Summary}

\begin{figure}
\begin{center}
\includegraphics[width=1.0\textwidth]{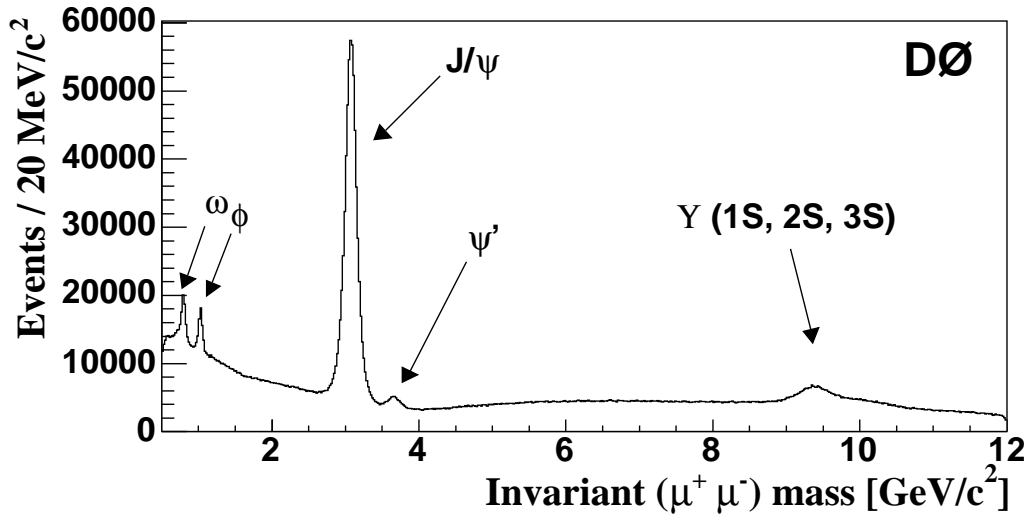}
\end{center}
\caption{Invariant mass of $\mu{^{+}}\mu{^{-}}$ pairs detected by the D$\O$ experiment in Run~II.}
\label{picmumu}
\end{figure}

This paper describes the upgraded D$\O$ muon system for Run~II of the Fermilab Tevatron collider. 
In the central region, a new A$\varphi$ trigger counter system is located inside the central 
iron toroid and the angular coverage of the counter system outside the iron toroids is improved. 
In the forward region, a new scintillation trigger counter system and a new mini drift 
tube tracking system are installed. All muon electronics is upgraded. The new shielding 
system is an important component of the muon upgrade that ensures that the D$\O$ muon system can 
operate efficiently in the background environment of a hadron collider. 
The channel counts for the Run~II D$\O$ muon system are given in Table~4.
Because of these upgrades,  
the muon identification and triggering capability is significantly enhanced, compared to that of the
D$\O$ Run~I muon detector. The performance of the D$\O$
Run~II muon detector during physics data collection demonstrates that design goals of 
the muon detector upgrade have been accomplished \cite{ref1,ref2,ref3}. The invariant mass distribution 
of dimuon pairs corresponding to an integrated luminosity of 0.20~fb$^{-1}$
is shown in Fig.~\ref{picmumu} and demonstrates the 
powerful triggering and identification capabilities of the Run~II D$\O$ muon system. 

\newpage

\begin{thebibliography}{99}

\bibitem{refabachi}
S.~Abachi, et al., Nucl. Instr. and Meth. A 338 (1994) 185. 

\bibitem{refbrown}
C.~Brown, et al., Nucl. Instr. and Meth. A 279 (1989) 331; \\
Yu.~Antipov, et al., Nucl. Instr. and Meth. A 297 (1990) 121.

\bibitem{refabramov}
V.~Abramov, et al., Nucl. Instr. and Meth. A 419 (1998) 660.

\bibitem{refyamada}
R.~Yamada, et al., IEEE Trans. Magnetics 28 (1) (1992) 520.

\bibitem{refvector}
Vector Fields Limited, 24 Bankside, Kidlington, Oxford, OX5 1JE, England.

\bibitem{refmokhov}
N.~V.~Mokhov, The MARS Code System User's Guide, Fermilab-FN-0628, 1995; \\
N.~V.~Mokhov, Status of MARS Code, Fermilab-Conf-03/053, 2003; \\
http://www-ap.fnal.gov/MARS/.

\bibitem{refbutler}
J.~M.~Butler, et al., Reduction of Tevatron and Main Ring Induced Backgrounds 
in the D$\O$ Detector, Fermilab-FN-0629, 1995. 

\bibitem{refdiehl}
H.~T.~Diehl, Aggressive Shielding Strategies for the Muon Upgrade, D$\O$ Note 2713, Fermilab, 1995; \\
V.~I.~Sirotenko, D.~S.~Denisov, H.~T.~Diehl, D$\O$ Shielding Optimization: Simulations and 
Measurements, D$\O$ Note 3417, Fermilab, 1998. 

\bibitem{refsirotenko}
N.~V.~Mokhov, V.~I.~Sirotenko, Possible Minimal Solutions for D$\O$ Detector Shielding 
Based on GCALOR and MARS13 Simulations, D$\O$ Note 2601, Fermilab, 1995.

\bibitem{refcosmo}
G.~Cosmo, S.~Giani, N.~J.~H{\o}imyr, et al., CERN-LHCC-95-70, CERN, Geneva, 
Switzerland, 1995.

\bibitem{refgaston}
A.~Bross, et al., The D$\O$ Scintillating Fiber Tracker, Proceedings of the SCIFI 97 Conference, 
p.~221, South Bend, Indiana, November, 1997. 

\bibitem{refacharya}
B.~S.~Acharya, et al., Nucl. Instr. and Meth. A 401 (1997) 45; \\ 
V.~Abramov, et al., IEEE  Trans. Nucl. Sci. NS-46 (4) (1999) 865.

\bibitem{refemi}
EMI 9902K, Thorn EMI Gencom Inc., 23 Madison Rd., Fairfield, NJ 07006.

\bibitem{refbicron}
Bicron Corporation, 12345 Kinsman Rd, Newbury, OH 44065-9677.

\bibitem{refevdoki}
V.~Evdokimov, Light Collection from Scintillation Counters using WLS Fibers and 
Bars, Proceedings of the SCIFI 97 Conference, p.~300, South Bend, Indiana, 
November, 1997.

\bibitem{refjoint}
MELZ, Electrozavodskaya str., 23, Moscow, Russia, 105023. 

\bibitem{refdupont}
Dupont De Nemours \& Co., 705 Canter Rd., Rt. 141, Wilmington, DE  19810-1025.

%
%
\bibitem{refbez}
V.~Bezzubov, et al., Fast Scintillation Counters with WLS Bars, 
Proceedings of the SCIFI 97 Conference, p.~210, South Bend, Indiana, November, 1997.

\bibitem{refkumarin}
S. Belikov, et al., Physical Characteristics of the SOFZ-105 Polymethyl
Methacrylate Secondary Emitter, Instruments and Experimental Technique, 36 (1993) p.~390.

\bibitem{refhv}
S.-C.~Ahn, et al., IEEE Nucl. Sci. Symposium, Vol.~2 (1991) 984.

\bibitem{refhanlet}
P.~Hanlet, et al., Nucl. Instr. and Meth. A 521 (2004) 343.

\bibitem{refglasteel}
Glasteel Tennessee, Inc., 175 Commerce Rd, Collierville, TN 38017.

\bibitem{refmarsh}
T.~Marshall, Nucl. Instr. and Meth. A 515 (2003) 50.

\bibitem{refbusza}
W.~Busza, Nucl. Instr. and Meth. A 265 (1988) 210.

\bibitem{refgarfield}
R.~Veenhof, GARFIELD, A Drift Chamber Simulation Program, CERN Program Library.
http://garfield.web.cern.ch/garfield/

\bibitem{refalexeev}
G.~D.~Alexeev, et al., Nucl. Instr. and Meth. A 473 (2001) 269.

\bibitem{refzhao}
T.~Zhao, et al., IEEE Trans. Nucl. Sci. NS-49 (3) (2002) 1092.

\bibitem{refbaldin}
B.~Baldin, et al., IEEE Trans. Nucl. Sci. NS-42 (4) (1995) 736.

\bibitem{refarai}
Y.~Arai, F.~Sudo and T.~Emura, Developments of time memory cell VLSI's, KEK
Preprint 93-49, 1993.

\bibitem{refmotorola}
http://gocct.com/products/vme/motorola680x0.htm

\bibitem{ref1}
V.~M.~Abazov, et al., Phys. Rev. Lett. 93 (2004) 141801.

\bibitem{ref2}
V.~M.~Abazov, et al., Phys. Rev. Lett. 93 (2004) 162002.

\bibitem{ref3}
V.~M.~Abazov, et al., hep-ex/0409043; Fermilab-Pub-04/225-E (2004).

\end{thebibliography}
\end{document}